%% file: Sohoma.tex
\documentclass[runningheads,a4paper]{llncs}

\usepackage{amssymb}
\usepackage{graphicx}
\usepackage[table]{xcolor}
\usepackage{amsmath,amsfonts}
\usepackage{epsfig}
\usepackage{subfigure}
\usepackage{hyperref}
\usepackage{pstricks}
\usepackage[english]{babel}
\usepackage{pstricks-add}
\usepackage[ruled,vlined]{algorithm2e}
\usepackage{multirow}

\usepackage{url}
\urldef{\mailsa}\path|{firstname.lastname}@cran.uhp-nancy.fr|
\newcommand{\keywords}[1]{\par\addvspace\baselineskip
\noindent\keywordname\enspace\ignorespaces#1}

\graphicspath{{./pictures/}}

\begin{document}

\mainmatter  % start of an individual contribution

\title{Key Factors for Information Dissemination on Communicating Products and Fixed Databases}

\titlerunning{Lecture Notes in Computer Science: Authors' Instructions}

\author{Sylvain KUBLER\and William DERIGENT\and Andr\'e THOMAS\and \'Eric RONDEAU\\}
\authorrunning{Lecture Notes in Computer Science: Authors' Instructions}

\institute{Research Centre for Automatic Control of Nancy, Nancy-University, CNRS,\\
Boulevard des Aiguillettes, F-54506 Vand\oe{}uvre-l\`es-Nancy, France\\
\mailsa\\
%\url{http://www.springer.com/lncs}
}


\toctitle{Lecture Notes in Computer Science}
\tocauthor{Authors' Instructions}
\maketitle

\begin{abstract}
Intelligent products carrying their own information are more and more present nowadays. In recent years, some authors argued the usage of such products for the Supply Chain Management Industry. Indeed, a multitude of informational vectors take place in such environments like fixed databases or manufactured products on which we are able to embed significant proportion of data. By considering distributed database systems, we can allocate specific data fragments to the product useful to manage its own evolution. The paper aims to analyze the Supply Chain performance according to different strategies of information distribution. Thus, different distribution patterns between informational vectors are studied. The purpose is to determine the key factors which lead to improve information distribution performance in term of time properties.
\end{abstract}
%\keywords{Distributed Database Systems}%, Information Dissemination, Supply Chain, Communicating product

\section{Introduction}\label{Sec1:Intro}
Intelligent products or products carrying their own information are more and more present nowadays. \cite{Ley2007} quotes the example of clothes able to carry their own information and thus enabling the washing machine to automatically adapt its washing program. In one of our previous works, \cite{KublerIMS2010} highlight several possible scenarios in different sectors: Supply Chain Management, healthcare \cite{Ausen2006}, home automation. Such applications rely on ever more complex information systems using a multitude of information vectors, in order to allow product information to be available anywhere and at anytime. These vectors may be fixed (desktop computers) or mobile devices (PDA, laptops, sensors, RFID technologies\ldots), short-lived or even invisibles (concept of disappearing computer, ubiquitous computing). More generally, the concept of the Internet of Things \cite{Gershenfeld2004} based on the RFID usage enables to access to information disseminated on any kind of physical object and to develop new smart services and applications.

According to Meyer \cite{Meyer2009}, in the context of supply chain management, few researches has been conducted on "intelligence at object", i.e products carrying their own information and intelligence. In fact, most of the time, products are only given an identifier (stored in a RFID tag) referring to a software agent or a database (approach used by \cite{Morel2003}). This mode of information management is diametrically opposed to works initiated since $2003$ by the PDMS (Product-Driven Manufacturing Systems) community, which advocates a physical information distribution on the product. In that case, a product carries physically a part, or even the totality of the information needed for its manufacturing or to manage its evolution all along its life cycle. Our previous work \cite{KublerIESM2011} aimed at prototyping a new type of materials, in which it is possible to write a significant quantity of information by inserting thousands of micro RFID tags. This new type of material is then referred to "communicating material". We developed an industrial process to produce a communicating textile with up to $1500tags/m^{2}$. Meyer concurs with the PDMS community by stressing the fact, in an increasingly interconnected and interdependent world involving many actors issued from different domains, supply chain information should not be stored in a single database but should be distributed all over the supply chain network. In fact, substantial information distribution improves data accessibility and availability, compared to centralized architectures. However, update mechanisms of the distributed information are needed in order to avoid problems related to data consistency and integrity. This type of architecture is thus more complex to design than centralized architectures.
As a result, product information can be spread out on mobile or fixed devices or even directly on the product, via simple RFID tags or communicating materials. Centralized architectures or highly distributed architectures can be employed. One might then wonder what the optimal information distribution is. The present paper aims to study the different ways to distribute information over a network composed of centralized, distributed databases and "communicating products", which may store information fragments as well. This study will determine the key factors which lead to improve information distribution performance. The performance is analyzed regarding the time required for accessing to the information system during the product life cycle.
Based on this influent factors determination, in a further work, we will be able to implement an experimental design leading us to control the best way to disseminate information on the informational vectors.

This question is addressed in several steps. First, the data distribution is introduced, and then an overview on conducted researches on distributed databases over fixed and mobile devices is presented in section~\ref{Sec2:DDBS}. Then, a case study extracted from this overview and adapted to our context is detailed in section~\ref{Sec3:StudyCase}. It only considers two types of informational vectors: fixed computers and communicating products. This case study is then used as a basis of comparison and evaluation between two different architectures of information distribution (one forbids data allocation on products while the other allows it). The evaluation process relies on several specific tools and a methodology using jointly two discrete-even simulators, one using Petri Nets (CPN tools) and the other dedicated to network protocol development and measurement (OPNET). This simulator looks for evaluating the manufacturing lead-time of a given number of communicating products all along the supply chain, by taking into account manufacturing run times, network delays, times to read/write information for both distributed databases and communicating products. This piece of software is presented in the section~\ref{Sec4:EvalSyst}. Finally, the section~\ref{Sec5:Result} presents the results obtained with the case study and an analysis of the main factors impacting on the performance of the information distribution.

\section{Distributed Database Systems}\label{Sec2:DDBS}

\subsection{General distribution framework}\label{Sec2:DDBS:General}
\begin{figure}[ht]
\psset{xunit=1mm,yunit=1mm,runit=1mm}
\begin{pspicture}(0,0)(122,27)
%\psframe(0,0)(122,27)
\centering

\put(76,19){\includegraphics[height=.8cm]{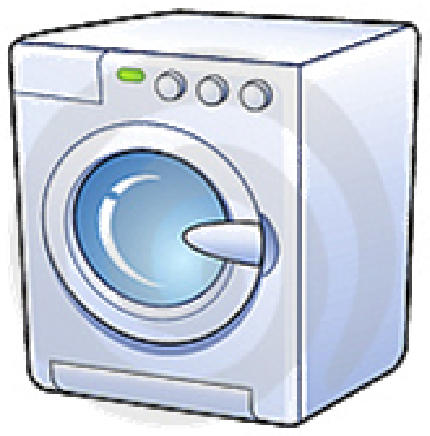}}
\pscurve[linecolor=red,linestyle=dashed](83.5,23)(85,21)(84,19)(91,20)

%\put(0,3){\includegraphics[width=90mm]{CoL.eps}}
%\put(58,19){\includegraphics[height=1.5cm]{clothe_MC.eps}}
\put(62,12){\includegraphics[height=1.5cm]{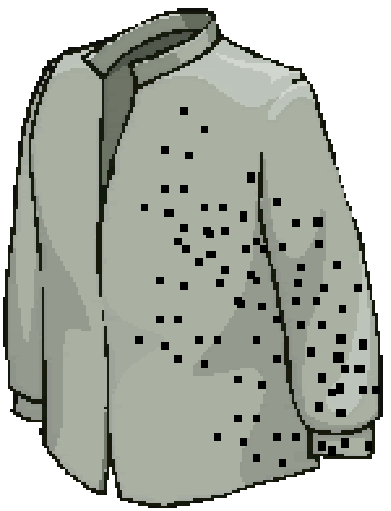}}
%\put(75,29){\includegraphics[height=.7cm]{wash_mach.eps}}

\put(5,12){\includegraphics[height=1.5cm]{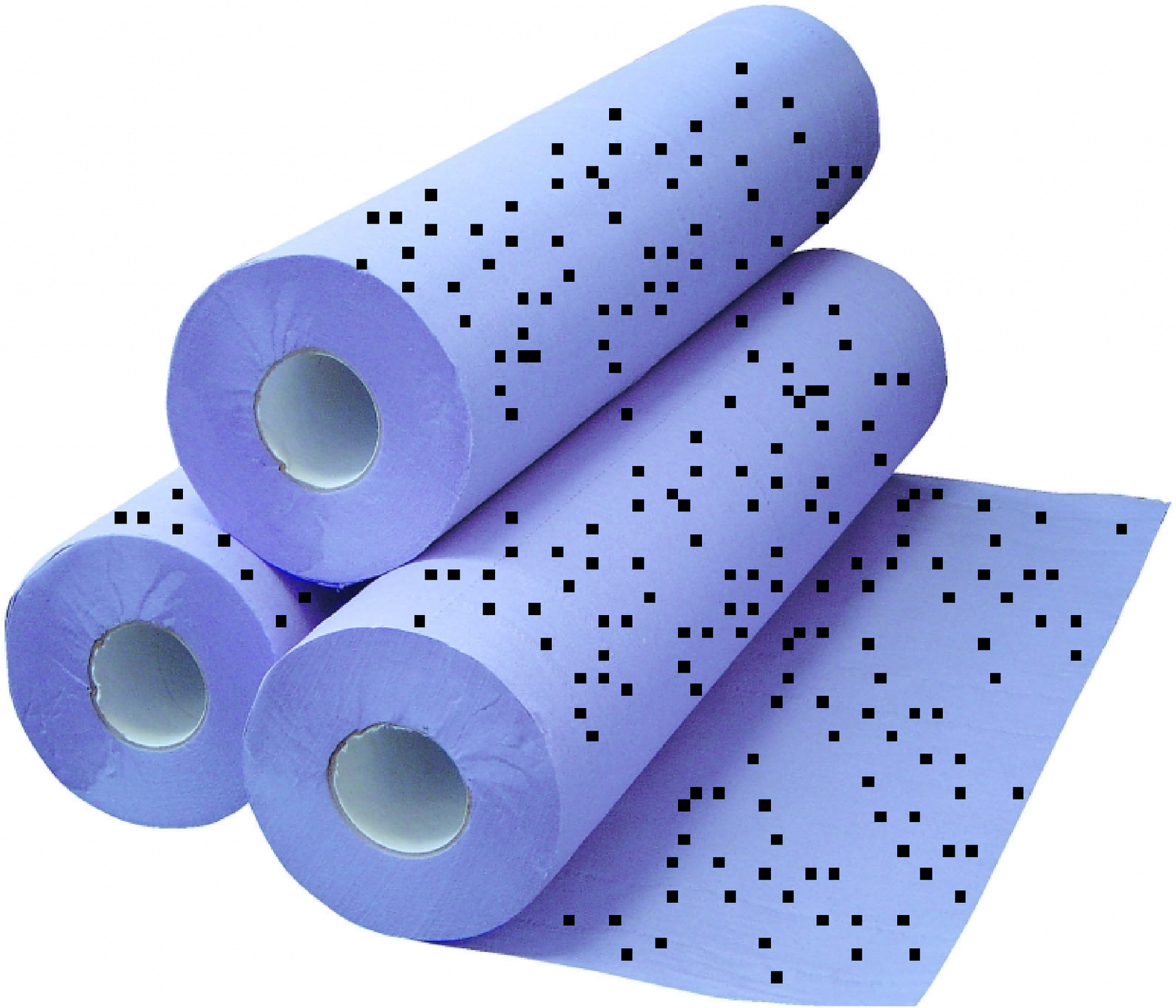}}
\put(107,12){\includegraphics[height=1.5cm]{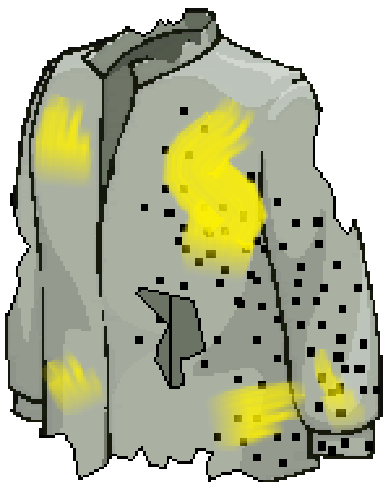}}
%\put(88,12){\includegraphics[height=.5cm]{recycle.eps}}
\put(88,17){\includegraphics[height=.9cm]{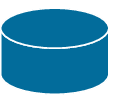}}
\rput(94,15.5){\scriptsize \emph{Fixed}}
\rput(94,13){\scriptsize \emph{database}}

\put(-20,7.75){
\psframe[fillstyle=solid,fillcolor=white,linewidth=.5pt,linecolor=red,linestyle=dashed](109,11)(112,13)
\rput(110.5,12){\tiny $F1$}
\put(3.5,-1.5){\psframe[fillstyle=solid,fillcolor=white,linewidth=.2pt](109,11)(112,13)
\rput(110.5,12){\tiny $F4$}}
\put(6.5,1){\psframe[fillstyle=solid,fillcolor=white,linewidth=.2pt](109,11)(112,13)
\rput(110.5,12){\tiny $F5$}}
}
\put(-94,2){\psframe[fillstyle=solid,fillcolor=white,linewidth=.2pt](109,11)(112,13)
\rput(110.5,12){\tiny $F2$}}
\put(-90.75,6){\psframe[fillstyle=solid,fillcolor=white,linewidth=.2pt](109,11)(112,13)
\rput(110.5,12){\tiny $F3$}}

\rput(30,8){Manufacturing parties}
\rput(82,8){Users}
\rput(110,8){Recyclers}

\pspolygon[fillstyle=solid,fillcolor=blue](0,0)(60,0)(63,2.5)(60,5)(0,5)(0,0)
\rput(31.5,2.5){\tiny \textcolor{white}{\textbf{Beginning of Life}}}
\pspolygon[fillstyle=solid,fillcolor=blue](61,0)(99,0)(102,2.5)(99,5)(61,5)(64,2.5)
\rput(83,2.5){\tiny \textcolor{white}{\textbf{Middle of Life}}}
\put(19,0){
\pspolygon[fillstyle=solid,fillcolor=blue](81,0)(100,0)(103,2.5)(100,5)(81,5)(84,2.5)
\rput(92.75,2.5){\tiny \textcolor{white}{\textbf{End of Life}}}}
\put(5,7){
\definecolor{bleubob}{rgb}{0.67,0.67,1.00}
\scalebox{.3}{
\put(55,3){

\put(35,6){
\psline[linestyle=dashed,linewidth=6pt]{->}(-30,27)(-10,27)
\psline[linestyle=dashed,linewidth=6pt]{->}(60,27)(80,27)

\psframe[fillstyle=solid,fillcolor=bleubob](0,8)(12,28)
\pscircle[linewidth=.05pt,fillstyle=solid,fillcolor=black](2,27){.5}
\pscircle[linewidth=.05pt,fillstyle=solid,fillcolor=black](2,21){.5}
\pscircle[linewidth=.05pt,fillstyle=solid,fillcolor=black](2,15){.5}
\pscircle[linewidth=.05pt,fillstyle=solid,fillcolor=black](2,9){.5}

%\pscircle[linewidth=.05pt,fillstyle=solid,fillcolor=lightgray](6,27){.75}
\pscircle[linewidth=.05pt,fillstyle=solid,fillcolor=black](6,24){.5}
\pscircle[linewidth=.05pt,fillstyle=solid,fillcolor=black](6,18){.5}
\pscircle[linewidth=.05pt,fillstyle=solid,fillcolor=black](6,13){.5}

\pscircle[linewidth=.05pt,fillstyle=solid,fillcolor=black](10,27){.5}
\pscircle[linewidth=.05pt,fillstyle=solid,fillcolor=black](10,21){.5}
\pscircle[linewidth=.05pt,fillstyle=solid,fillcolor=black](10,15){.5}
\pscircle[linewidth=.05pt,fillstyle=solid,fillcolor=black](10,9){.5}
}

\put(55,6){
\psframe[fillstyle=solid,fillcolor=bleubob](0,8)(12,28)
\pscircle[linewidth=.05pt,fillstyle=solid,fillcolor=black](2,27){.5}
\pscircle[linewidth=.05pt,fillstyle=solid,fillcolor=black](2,21){.5}
\pscircle[linewidth=.05pt,fillstyle=solid,fillcolor=black](2,15){.5}
\pscircle[linewidth=.05pt,fillstyle=solid,fillcolor=black](2,9){.5}

%\pscircle[linewidth=.05pt,fillstyle=solid,fillcolor=lightgray](6,27){.75}
\pscircle[linewidth=.05pt,fillstyle=solid,fillcolor=black](6,24){.5}
\pscircle[linewidth=.05pt,fillstyle=solid,fillcolor=black](6,18){.5}
\pscircle[linewidth=.05pt,fillstyle=solid,fillcolor=black](6,13){.5}

\pscircle[linewidth=.05pt,fillstyle=solid,fillcolor=black](10,27){.5}
\pscircle[linewidth=.05pt,fillstyle=solid,fillcolor=black](10,21){.5}
\pscircle[linewidth=.05pt,fillstyle=solid,fillcolor=black](10,15){.5}
\pscircle[linewidth=.05pt,fillstyle=solid,fillcolor=black](10,9){.5}
}

\put(75,6){
\psframe[fillstyle=solid,fillcolor=bleubob](0,8)(12,28)
\pscircle[linewidth=.05pt,fillstyle=solid,fillcolor=black](2,27){.5}
\pscircle[linewidth=.05pt,fillstyle=solid,fillcolor=black](2,21){.5}
\pscircle[linewidth=.05pt,fillstyle=solid,fillcolor=black](2,15){.5}
\pscircle[linewidth=.05pt,fillstyle=solid,fillcolor=black](2,9){.5}

%\pscircle[linewidth=.05pt,fillstyle=solid,fillcolor=lightgray](6,27){.75}
\pscircle[linewidth=.05pt,fillstyle=solid,fillcolor=black](6,24){.5}
\pscircle[linewidth=.05pt,fillstyle=solid,fillcolor=black](6,18){.5}
\pscircle[linewidth=.05pt,fillstyle=solid,fillcolor=black](6,13){.5}

\pscircle[linewidth=.05pt,fillstyle=solid,fillcolor=black](10,27){.5}
\pscircle[linewidth=.05pt,fillstyle=solid,fillcolor=black](10,21){.5}
\pscircle[linewidth=.05pt,fillstyle=solid,fillcolor=black](10,15){.5}
\pscircle[linewidth=.05pt,fillstyle=solid,fillcolor=black](10,9){.5}
}

\put(35,30){
\psframe[fillstyle=solid,fillcolor=bleubob](0,8)(12,28)
\pscircle[linewidth=.05pt,fillstyle=solid,fillcolor=black](2,27){.5}
\pscircle[linewidth=.05pt,fillstyle=solid,fillcolor=black](2,21){.5}
\pscircle[linewidth=.05pt,fillstyle=solid,fillcolor=black](2,15){.5}
\pscircle[linewidth=.05pt,fillstyle=solid,fillcolor=black](2,9){.5}

%\pscircle[linewidth=.05pt,fillstyle=solid,fillcolor=lightgray](6,27){.75}
\pscircle[linewidth=.05pt,fillstyle=solid,fillcolor=black](6,24){.5}
\pscircle[linewidth=.05pt,fillstyle=solid,fillcolor=black](6,18){.5}
\pscircle[linewidth=.05pt,fillstyle=solid,fillcolor=black](6,13){.5}

\pscircle[linewidth=.05pt,fillstyle=solid,fillcolor=black](10,27){.5}
\pscircle[linewidth=.05pt,fillstyle=solid,fillcolor=black](10,21){.5}
\pscircle[linewidth=.05pt,fillstyle=solid,fillcolor=black](10,15){.5}
\pscircle[linewidth=.05pt,fillstyle=solid,fillcolor=black](10,9){.5}
}

\put(55,30){
\psframe[fillstyle=solid,fillcolor=bleubob](0,8)(12,28)
\pscircle[linewidth=.05pt,fillstyle=solid,fillcolor=black](2,27){.5}
\pscircle[linewidth=.05pt,fillstyle=solid,fillcolor=black](2,21){.5}
\pscircle[linewidth=.05pt,fillstyle=solid,fillcolor=black](2,15){.5}
\pscircle[linewidth=.05pt,fillstyle=solid,fillcolor=black](2,9){.5}

%\pscircle[linewidth=.05pt,fillstyle=solid,fillcolor=lightgray](6,27){.75}
\pscircle[linewidth=.05pt,fillstyle=solid,fillcolor=black](6,24){.5}
\pscircle[linewidth=.05pt,fillstyle=solid,fillcolor=black](6,18){.5}
\pscircle[linewidth=.05pt,fillstyle=solid,fillcolor=black](6,13){.5}

\pscircle[linewidth=.05pt,fillstyle=solid,fillcolor=black](10,27){.5}
\pscircle[linewidth=.05pt,fillstyle=solid,fillcolor=black](10,21){.5}
\pscircle[linewidth=.05pt,fillstyle=solid,fillcolor=black](10,15){.5}
\pscircle[linewidth=.05pt,fillstyle=solid,fillcolor=black](10,9){.5}
}

\put(75,30){
\psframe[fillstyle=solid,fillcolor=bleubob](0,8)(12,28)
\pscircle[linewidth=.05pt,fillstyle=solid,fillcolor=black](2,27){.5}
\pscircle[linewidth=.05pt,fillstyle=solid,fillcolor=black](2,21){.5}
\pscircle[linewidth=.05pt,fillstyle=solid,fillcolor=black](2,15){.5}
\pscircle[linewidth=.05pt,fillstyle=solid,fillcolor=black](2,9){.5}

%\pscircle[linewidth=.05pt,fillstyle=solid,fillcolor=lightgray](6,27){.75}
\pscircle[linewidth=.05pt,fillstyle=solid,fillcolor=black](6,24){.5}
\pscircle[linewidth=.05pt,fillstyle=solid,fillcolor=black](6,18){.5}
\pscircle[linewidth=.05pt,fillstyle=solid,fillcolor=black](6,13){.5}

\pscircle[linewidth=.05pt,fillstyle=solid,fillcolor=black](10,27){.5}
\pscircle[linewidth=.05pt,fillstyle=solid,fillcolor=black](10,21){.5}
\pscircle[linewidth=.05pt,fillstyle=solid,fillcolor=black](10,15){.5}
\pscircle[linewidth=.05pt,fillstyle=solid,fillcolor=black](10,9){.5}
}
}}

\put(-82,3){\psframe[fillstyle=solid,fillcolor=white,linewidth=.2pt,linecolor=gray](109,11)(112,13)
\rput(110.5,12){\tiny $F3$}}
\put(-70,-5){\psframe[fillstyle=solid,fillcolor=white,linewidth=.2pt,linecolor=gray](109,11)(112,13)
\rput(110.5,12){\tiny $F2$}}
\put(-75.6,-3){\psframe[fillstyle=solid,fillcolor=white,linewidth=.2pt,linecolor=gray](109,11)(112,13)
\rput(110.5,12){\tiny $F2$}}
\put(-75.8,2){\psframe[fillstyle=solid,fillcolor=white,linewidth=.2pt,linecolor=gray](109,11)(112,13)
\rput(110.5,12){\tiny $F3$}}

\put(-50,-3){\psframe[fillstyle=solid,fillcolor=white,linewidth=.5pt,linecolor=red,linestyle=dashed](109,11)(112,13)
\rput(110.5,12){\tiny $F2$}}
\put(-49,2){\psframe[fillstyle=solid,fillcolor=white,linewidth=.2pt,linecolor=gray](109,11)(112,13)
\rput(110.5,12){\tiny $F3$}}

\put(45,0){
%\put(-50,-3){\psframe[fillstyle=solid,fillcolor=white,linewidth=.5pt,linecolor=red,linestyle=dashed](109,11)(112,13)
%\rput(110.5,12){\tiny F2}}
\put(-49,2){\psframe[fillstyle=solid,fillcolor=white,linewidth=.2pt,linecolor=gray](109,11)(112,13)
\rput(110.5,12){\tiny $F3$}}}

\put(47,-2){
\psarc[linecolor=red,linestyle=dashed]{-}(22,19){8}{180}{270}
\psarc[linecolor=red,linestyle=dashed]{-}(22.5,20){5}{180}{270}
\psarc[linecolor=red,linestyle=dashed]{-}(22.5,20.25){1.5}{180}{270}
}

}

\end{pspicture}
\caption{Information distribution on products and fixed databases}\label{Fig:CoL}
\end{figure}

During the product lifecycle, users could have to access to product information for diverse reasons, either in the design phase, the usage phase or still the recycle phase. As exposed before, information can be stored both on the product and-or on fixed databases. Information are therefore bind to one or more relational data models, which have to be fragmented and distributed by the best way on these informational vectors. One example retracing briefly a bobbin lifecycle is presented in the figure~\ref{Fig:CoL}. We can see $5$ data fragments [$F1$..$F5$] distributed between the product and the database ($F1$, $F4$, $F5$ allocated to the database system and $F2$, $F3$ to the product). By reconsidering the example given by \cite{Ley2007}, the washing machine could access to data fragments located both on the product and on the database according to its queries. In our researches, we are looking for assessing different distribution patterns between both informational vectors (manufactured products and fixed databases) by taking into account the access times for reaching information. Work on distributed databases considering fixed and mobile environments are introduced in the next section.

\vspace{-1cm}

\subsection{Distributed databases through literature}\label{Sec2:DDBS:Literature}
The main objective of the data dissemination in an information system is to make the dissemination process transparent for users: location, partitioning and replication transparency. Indeed, no matter why, where and how the data repartition is achieved from a user's point of view. The structured data distribution regarding relational data models (within distributed database systems framework) is carried out in two steps: The partitioning of the data model follows up by the allocation phase of the resulting fragments. Many approaches and mechanisms exist for ensuring the best partitioning and allocation of the relational model regarding to the environment and some applicative constraints.

Basically, the partitioning aim is to subdivide the concerned relational data model. Thus, the resulting fragments will be allocated at specific informational vectors in order to improve system performance. Three types of fragmentation exist: vertical \cite{Navathe1984}, horizontal \cite{Apers1988}, mixed/hybrid \cite{Navathe1995}. %In the following we will introduce them briefly.
The vertical fragmentation aims to break up a relation into a set of relations. It consists in dividing the attributes of a relation (i.e the columns of a relational table). The horizontal fragmentation aims to break the large number of object instances into disjoint subsets. It consists in partitioning the tuples of a relation (i.e the rows of a relational table). The hybrid fragmentation first divides the relation horizontally, and then splits each of the obtained fragments vertically or vice versa. %Usually, a vertical fragmentation is used during the design phase (static phase) of the database contrary to a horizontal fragmentation more suitable for dynamic systems. A static fragmentation or allocation is not suitable if the applicative characteristics evolve strongly (number of requests\ldots) since the result of these operations is never called into question, which may be an issue. The dynamic approaches overcome this problem but they need more memory resources and generate overload traffic on the network (necessary for the proper functioning of the algorithms). It is therefore important to define the applicative expectations in order to adopt the suitable approach.

As stated previously, the allocation phase takes place subsequently to the fragmentation phase and the aim is to establish the optimal fragment assignation on the databases. Usually, methods tend to assign fragments to the clients requesting them mostly via objective functions that we attempt to minimize or maximize \cite{Huang2001,Hababeh2005}. Let us note that it is possible to perform data replications, or in other words, to replicate a same fragment on several databases. This has the dual benefit of maintaining the system reliability and of increasing performance (e.g reduction of the overload traffic) \cite{Padmanabhan2008}. However, replication mechanisms are necessary for handling both the modification broadcast (updates) on replica and also the information access rights (to authorize one site or one group of sites to modify replica). The applicative expectations have an influence on the mechanism to implement and actually two parameters have to be characterized: \emph{When} and \emph{Where}? \emph{When} do the updates have to be propagated? Two modes are available: Synchronous (S) and Asynchronous (As). The As mode makes it possible to carry out local modification without needing to inform its peers (contrarily to the S mode). \emph{Where} do the updates have to be performed ? Two principles exist: Update everywhere (Ue) and Primary copy (Pc). The Pc principle allows one site to perform modifications on a data fragment contrarily to the Ue mode which allows one group of sites. Finally, four types of replication may be considered: Ue-S, Ue-As, Pc-S and PC-As. %\cite{Lubinski2000} introduce a tripartite model reflecting the incontrovertible criteria for choosing the suitable protocol, namely: the data consistency, the data availability and minimizing communication costs. They highlight the fact that it is impossible to satisfy the three objectives simultaneously and as a result, a compromise must be found.
Also note that the memory storage limitation of mobile devices is a problem frequently encountered in the literature. Accordingly, some authors focus on the data summarization \cite{Chan2003,Lubinski2000a} (subclass of the data mining) whose primary aim is to reduce the information somehow. \cite{Chan2005} list the summarization methods used for distributed database systems and mention the fragmentation/allocation method (used in our study).

A multitude of interesting approaches are proposed in the literature, we therefore feel it is necessary to confront our proposition with them in order to compare and assess our distribution models. In this sense, works reported by Hababeh \cite{Hababeh2005} seem interesting as basis of comparison. Indeed, a fragment distribution method is developed and then applied on a case study, which can be easily extended to our application. In what follows, two distribution architectures will be defined, the first one does not consider the presence of communicating products able to store data fragments, i.e all information is located on databases. In fact, we rely on the distribution defined by Hababeh. The second one considers communicating products able to store data fragments, thus, diverse distribution patterns of fragments between the product and databases will be possible. %Let us note that the databases are also distributed in line with the distribution schema of Hababeh.
The next section introduces this case study and then the adaptation realized in this paper.

\section{Case study presentation}\label{Sec3:StudyCase}
\subsection{Reference distribution pattern}\label{Sec3:StudyCase:SchemaRef}
\begin{figure}[ht]
\centering
\psset{xunit=1mm,yunit=1mm,runit=1mm}
\begin{pspicture}(0,0)(122,47)
%\psframe(0,0)(122,47)

\definecolor{jaune}{rgb}{1.00,1.00,0.78}
\definecolor{jaunef}{rgb}{1.00,1.00,0.50}

%------ Cluster 1
\put(30.5,20){
\rput(30.5,26){\scriptsize Allocated fragments:}
\rput(30.5,23.25){\scriptsize $F1$, $F5$, $F8$}
\rput(30.6,20.5){\scriptsize \textcolor{gray}{\emph{Cluster-Location 1}}}

\psellipse[linecolor=gray](30.5,11)(15,8)
%bus
\psline(20,7)(41,7)
\psline(30.5,7)(30.5,-2)

\psline(24.5,13)(24.5,7)
\rput(24.5,14){\includegraphics[width=5mm]{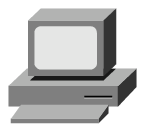}}
\psline(34.75,13)(34.75,7)
\rput(34.75,14){\includegraphics[width=5mm]{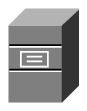}}
\rput(20,11){\tiny client 1}
\rput(41,10){\tiny DB 1}
\pscurve[linestyle=dashed,linecolor=lightgray,linewidth=.5pt](24.5,14.5)(6,19)(-4,15.5)
\rput(18.6,9){\tiny \textcolor{red}{machine 1}}

\put(-28,-20){
\pspolygon[fillstyle=solid,fillcolor=jaune,linecolor=gray](24,30)(24,45)(38.3,45)(41,42)(41,30)(24,30)
\pspolygon[fillstyle=solid,fillcolor=jaunef,linecolor=gray](38.3,45)(41,42)(38.3,42)
\psframe[linestyle=none,fillstyle=solid,fillcolor=white](38,45.001)(38.6,46)
\psframe[linestyle=none,fillstyle=solid,fillcolor=white](41.001,41)(42,43)
\put(-3.6,0){
\put(28,43){\tiny \underline{Read}}
\put(28,40){\tiny $F1$\emph{(540)}}
\put(28,38){\tiny $F4$\emph{(90)}}
\put(28,36){\tiny $F5$\emph{(54)}}
\put(28,34){\tiny $F8$\emph{(720)}}
\put(37,43){\tiny \underline{Upd}}
\put(37,40){\tiny $F1$\emph{(286)}}
\put(37,38){\tiny $F3$\emph{(110)}}
\put(37,36){\tiny $F4$\emph{(220)}}
\put(37,34){\tiny $F5$\emph{(110)}}
\put(37,32){\tiny $F8$\emph{(220)}}
}
}
}

%-------Cluster 2
\put(-10,-3){
\rput(30.5,25.5){\scriptsize Allocated fragments:}
\rput(30.5,23){\scriptsize $F4$, $F7$, $F8$}
\rput(30.5,20.5){\scriptsize \textcolor{gray}{\emph{Cluster-Location 2}}}

\psellipse[linecolor=gray](30.5,11)(15,8)
\psline(20,15)(60,15)

\psline(24.5,9)(24.5,15)
\rput(24.5,8){\includegraphics[width=5mm]{pc.eps}}
\psline(34.75,9)(34.75,15)
\rput(34.75,8){\includegraphics[width=5mm]{file_server.eps}}
\rput(20,12){\tiny client 2}
\rput(41,12){\tiny DB 2}
\pscurve[linestyle=dashed,linecolor=lightgray,linewidth=.5pt](24.5,8.5)(12,20)(13,29)
\rput(17,10){\tiny \textcolor{red}{machine 2}}

\put(-12,-1){
\pspolygon[fillstyle=solid,fillcolor=jaune,linecolor=gray](24,30)(24,45)(38.3,45)(41,42)(41,30)(24,30)
\pspolygon[fillstyle=solid,fillcolor=jaunef,linecolor=gray](38.3,45)(41,42)(38.3,42)
\psframe[linestyle=none,fillstyle=solid,fillcolor=white](38,45.001)(38.6,46)
\psframe[linestyle=none,fillstyle=solid,fillcolor=white](41.001,41)(42,43)
\put(-3.6,0){
\put(28,43){\tiny \underline{Read}}
\put(28,40){\tiny $F2$\emph{(180)}}
\put(28,38){\tiny $F4$\emph{(45)}}
\put(28,36){\tiny $F5$\emph{(180)}}
\put(28,34){\tiny $F7$\emph{(315)}}
\put(28,32){\tiny $F8$\emph{(540)}}
\put(37,43){\tiny \underline{Upd}}
\put(37,40){\tiny $F2$\emph{(66)}}
\put(37,38){\tiny $F4$\emph{(132)}}
\put(37,36){\tiny $F5$\emph{(110)}}
\put(37,34){\tiny $F7$\emph{(110)}}
\put(37,32){\tiny $F8$\emph{(110)}}
}
}
}

\put(71,-3){
\rput(30.5,26){\scriptsize Allocated fragments:}
\rput(30.5,23.25){\scriptsize $F2$, $F3$, $F5$, $F6$, $F7$}
\rput(30.5,20.5){\scriptsize \textcolor{gray}{\emph{Cluster-Location 3}}}

\psellipse[linecolor=gray](30.5,11)(15,8)
\psline(-20,15)(41,15)

\psline(24.5,9)(24.5,15)
\rput(24.5,8){\includegraphics[width=5mm]{pc.eps}}
\psline(34.75,9)(34.75,15)
\rput(34.75,8){\includegraphics[width=5mm]{file_server.eps}}
\rput(20,12){\tiny client 3}
\rput(41,12){\tiny DB 3}
\pscurve[linestyle=dashed,linecolor=lightgray,linewidth=.5pt](24.5,8.7)(12,20)(15,30)
\rput(17,10){\tiny \textcolor{red}{machine 3}}

\put(-10,2){
\pspolygon[fillstyle=solid,fillcolor=jaune,linecolor=gray](24,28)(24,45)(38.3,45)(41,42)(41,28)(24,28)
\pspolygon[fillstyle=solid,fillcolor=jaunef,linecolor=gray](38.3,45)(41,42)(38.3,42)
\psframe[linestyle=none,fillstyle=solid,fillcolor=white](38,45.001)(38.6,46)
\psframe[linestyle=none,fillstyle=solid,fillcolor=white](41.001,41)(42,43)
\put(-3.6,0){
\put(28,43){\tiny \underline{Read}}
\put(28,40){\tiny $F1$\emph{(315)}}
\put(28,38){\tiny $F2$\emph{(45)}}
\put(28,36){\tiny $F3$\emph{(360)}}
\put(28,34){\tiny $F5$\emph{(315)}}
\put(28,32){\tiny $F6$\emph{(225)}}
\put(28,30){\tiny $F7$\emph{(90)}}
\put(37,43){\tiny \underline{Upd}}
\put(37,40){\tiny $F1$\emph{(55)}}
\put(37,38){\tiny $F2$\emph{(330)}}
\put(37,36){\tiny $F3$\emph{(330)}}
\put(37,34){\tiny $F5$\emph{(110)}}
\put(37,32){\tiny $F6$\emph{(55)}}
\put(37,30){\tiny $F8$\emph{(220)}}
}
}
}

\rput(61,10){\includegraphics[width=27mm]{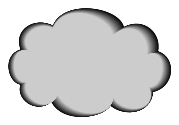}}
\rput(61,10){\textbf{Network}}

\end{pspicture}
\caption{Optimal distribution architecture established by \cite{Hababeh2005}}\label{Fig:arch_phys}
\end{figure}

Hababeh proposes a fragment distribution approach based on a two step process: first, the sites (clients and databases) are clustered according to communication costs, and then data fragments are allocated to the different clusters via an optimization function. %This function relies on several criteria such as communication costs, storage costs, query patterns\ldots
This approach is applied on a specific case study, including $3$ databases, $3$ clients which perform read and write accesses on a set of fragments ($8$ in total: [$F1$..$F8$]). The resulting optimal allocation \cite{Hababeh2005} is depicted on the figure~\ref{Fig:arch_phys}, the access pattern to the $8$ fragments performed by each client is specified, too (number in brackets indicates the number of bytes).
The next section formulates the adaptation of this case study to our logistic scenario. In fact, we match parameters and data defined  by Hababeh with the supply chain tasks, actors: number of databases and clients, query patterns, data fragments\ldots

\vspace{-1cm}

\subsection{Adaptation of the logistic process}\label{Sec3:StudyCase:adaptation}
A supply chain process consists of a set of tasks in a planned pattern or sequence (rout sheet). These tasks may correspond to manufacture operations, transport phases\ldots and can be performed on different physical locations (e.g supplier). These locations may dispose of local databases and can implement their own information systems (related to their tasks), but they can also access to remote databases if a collaboration between actors exists. As a matter of fact, databases are distributed (or federated) through one or more relational data models. Inspired by our current researches, the applicative framework considered is related to a supply chain management process dedicated to the textile industry.

In order to adapt the case study described previously to our logistic scenario, we consider three remote locations corresponding to the clusters $1$, $2$ and $3$ introduced on the figure~\ref{Fig:arch_phys}. One physical transformation is carried out on each location:
the operation $1$ and the operation $2$ (performed on the location $1$ and $2$ respectively) consist of cutting a set of \emph{bobbins $1$} and \emph{bobbins $2$} respectively, the operation $3$ (performed on the location $3$) consists of sewing the textile pieces resulting from the two previous operations.
%\begin{itemize}
%\item \emph{operation $1$ (location $1$)}: bobbin $1$ is cut on the location $1$,
%\item \emph{operation $2$ (location $2$)}: bobbin $2$ is cut on the location $2$,
%\item \emph{operation $3$ (location $3$)}: both textile pieces resulting from the precedent tasks (from the bobbin $1$ and $2$) are are assembled/sewed on the location $3$.
%\end{itemize}
Each location has one machine to achieve its own operation, this machine requires information after the arrival of products (range of product,  production order\ldots) and updates some of this information (notifications\ldots). Therefore, the applicative characteristics defined for each client in Hababeh will be matched to each machine (located in each cluster). In other words, the machine $2$ will have the same fragment access pattern (read and write accesses) than the client $2$ defined in Hababeh and so on. Likewise, each location disposes of a local database and shares a relational data model, this data model has to be distributed on the three databases. Taking into account of the input parameters defined in Hababeh (query pattern, architecture\ldots), the optimal distribution considered in our paper may be defined as shown in the figure~\ref{Fig:arch_phys}: $F1$,$F5$,$F8$ allocated to DB1, $F4$,$F7$,$F8$ to DB2\ldots% and $F2$,$F3$,$F5$,$F5$,$F7$ to DB3.

\begin{figure}[ht]
\centering
\scalebox{.85}{
\psset{xunit=1mm,yunit=1mm,runit=1mm}
\begin{pspicture}(0,15)(122,84)
%\psframe(0,15)(122,84)

\definecolor{vert}{rgb}{0.46,1.00,0.46}
\definecolor{bleu}{rgb}{0.56,0.95,0.99}
\definecolor{bleuc}{rgb}{0.84,0.98,0.99}
\definecolor{rougec}{rgb}{1.00,0.88,0.88}

\psframe[linewidth=.5pt,linecolor=gray,linestyle=dashed](89,15)(123,21)
\rput(106,18){\scriptsize $n,m$: number of products}
\put(-7.5,0){
%%%----------------Architecture: ODA ------
\rput(66,82){\textcolor{red}{\textbf{\underline{Architecture: ODA}}}}
\put(5,0){\scriptsize{

%%-----------------ligne fabrication : Decoupe1

\put(-12,0){
\put(15.5,74.5){$n$}
\psellipse(35,75)(7,3.5)
\rput(35,75){Bobbin 1}
\psline[linewidth=.7pt,arrowsize=0.12cm,arrowlength=0.6,arrowinset=0,linestyle=dashed]{->}(20,75)(28,75)
\put(40,70.5){PA}
\psline[linewidth=.7pt,arrowsize=0.12cm,arrowlength=0.6,arrowinset=0]{->}(42,75)(55.25,75)
\put(53,20){
\psframe[linewidth=.7pt](2.25,52.5)(15.75,57.5)
\psframe[linewidth=.7pt](3,53.25)(15,56.75)
\rput(9,55){cutting 1}
\psframe[fillstyle=solid,fillcolor=bleu,linewidth=.2pt](4,51.5)(14,53.5)
\rput(9,52.5){\tiny {cutting 1}}
}
}
\psline[linewidth=.7pt,arrowsize=0.12cm,arrowlength=0.6,arrowinset=0]{->}(56.6,75)(63,75)
\psline[linewidth=.7pt,arrowsize=0.12cm,arrowlength=0.6,arrowinset=0]{-}(77,75)(82,75)
\psline[linewidth=.7pt,arrowsize=0.12cm,arrowlength=0.6,arrowinset=0]{->}(82,75)(82,67.5)
\psellipse(70,75)(7,3.5)
\rput(70,75){End pt1}
\put(75,70.5){PA}

\put(18,74.5){
\tiny{
\put(-12,2.3){\psframe[shadow=true,fillstyle=solid,fillcolor=vert,linewidth=.5pt](2,2)(16,5)
\put(2.5,3){8`bob1@0.0}}
\pscircle[fillstyle=solid,fillcolor=green,linewidth=.5pt](1,3.5){1.2}
\rput(1,3.5){8}
}}

%%------------ligne fabrication : Decoupe2

\put(0,-20){\put(-12,0){
\put(15.5,74.5){$m$}
\psellipse(35,75)(7,3.5)
\rput(35,75){Bobbin 2}
\psline[linewidth=.7pt,arrowsize=0.12cm,arrowlength=0.6,arrowinset=0,linestyle=dashed]{->}(20,75)(28,75)
\put(40,70.5){PA}
\psline[linewidth=.7pt,arrowsize=0.12cm,arrowlength=0.6,arrowinset=0]{->}(42,75)(55.25,75)
\put(53,20){
\psframe[linewidth=.7pt](2.25,52.5)(15.75,57.5)
\psframe[linewidth=.7pt](3,53.25)(15,56.75)
\rput(9,55){cutting 2}
\psframe[fillstyle=solid,fillcolor=bleu,linewidth=.2pt](4,51.5)(14,53.5)
\rput(9,52.5){\tiny {cutting 2}}
}
}
\psline[linewidth=.7pt,arrowsize=0.12cm,arrowlength=0.6,arrowinset=0]{->}(56.6,75)(63,75)
\psline[linewidth=.7pt,arrowsize=0.12cm,arrowlength=0.6,arrowinset=0]{-}(77,75)(82,75)
\psline[linewidth=.7pt,arrowsize=0.12cm,arrowlength=0.6,arrowinset=0]{->}(82,75)(82,81.5)
\psellipse(70,75)(7,3.5)
\rput(70,75){End pt2}
\put(75,70.5){PA}

\put(18,74.5){
\tiny{
\put(-12,2.3){\psframe[shadow=true,fillstyle=solid,fillcolor=vert,linewidth=.5pt](2,2)(16,5)
\put(2.5,3){5`bob2@0.0}}
\pscircle[fillstyle=solid,fillcolor=green,linewidth=.5pt](1,3.5){1.2}
\rput(1,3.5){8}
}}
}

%%--------------ligne DB + couture

%%%%%--DB
\put(30.5,24){
\rput(19.5,41){\psovalbox{DB}}
\put(23,37.5){DB}
\psline[linewidth=.7pt,arrowsize=0.12cm,arrowlength=0.6,arrowinset=0]{<->}(19.5,38.5)(19.5,33.5)
\psline[linewidth=.7pt,arrowsize=0.12cm,arrowlength=0.6,arrowinset=0]{<->}(19.5,43.5)(19.5,47.5)
\psline[linewidth=.7pt,arrowsize=0.12cm,arrowlength=0.6,arrowinset=0]{<->}(23.6,41)(44.8,41)
%%--token
\put(14,37.5){\tiny{
\put(-20,0){
\psframe[shadow=true,fillstyle=solid,fillcolor=vert,linewidth=.5pt](2,0)(20,7)
\put(2.5,5){1`DB1@0.0+++}
\put(2.5,3){1`DB2@0.0+++}
\put(2.5,1){1`DB3@0.0}
}
\pscircle[fillstyle=solid,fillcolor=green,linewidth=.5pt](1,3.5){1.2}
\rput(1,3.5){3}
}}

\put(42.5,-14){
\psframe[linewidth=.7pt](2.25,52.5)(15.75,57.5)
\psframe[linewidth=.7pt](3,53.25)(15,56.75)
\rput(9,55){sewing}
\psframe[fillstyle=solid,fillcolor=bleu,linewidth=.2pt](4,51.5)(14,53.5)
\rput(9,52.5){\tiny {sewing}}
\psline[linewidth=.7pt,arrowsize=0.12cm,arrowlength=0.6,arrowinset=0]{->}(15.75,55)(24.2,55)
}

%%%%%--sewing
\put(55,0){
\psellipse(19.5,41)(8,3.5)
\rput(19.5,41){\scriptsize headdress}
\put(25.8,37){PA}
\psline[linewidth=.7pt,arrowsize=0.12cm,arrowlength=0.6,arrowinset=0,linestyle=dashed]{->}(28,41)(35.8,41)
}
}

}}

%%%----------------Architecture: ODAP ------
\put(0,-35){
\rput(66,82){\textcolor{red}{\textbf{\underline{Architecture: ODAP}}}}
\put(5,0){\scriptsize{

%%-----------------ligne fabrication : Decoupe1

\put(-12,0){
\put(15.5,74.5){$n$}
\psellipse(35,75)(7,3.5)
\rput(35,75){Bobbin 1c}
\psline[linewidth=.7pt,arrowsize=0.12cm,arrowlength=0.6,arrowinset=0,linestyle=dashed]{->}(20,75)(28,75)
\put(40,70.5){PAc}
\psline[linewidth=.7pt,arrowsize=0.12cm,arrowlength=0.6,arrowinset=0]{->}(42,75)(55.25,75)
\put(53,20){
\psframe[linewidth=.7pt](2.25,52.5)(15.75,57.5)
\psframe[linewidth=.7pt](3,53.25)(15,56.75)
\rput(9,55){cutting 1}
\psframe[fillstyle=solid,fillcolor=bleu,linewidth=.2pt](4,51.5)(14,53.5)
\rput(9,52.5){\tiny {cutting 1}}
}
}
\psline[linewidth=.7pt,arrowsize=0.12cm,arrowlength=0.6,arrowinset=0]{->}(56.6,75)(63,75)
\psline[linewidth=.7pt,arrowsize=0.12cm,arrowlength=0.6,arrowinset=0]{-}(77,75)(82,75)
\psline[linewidth=.7pt,arrowsize=0.12cm,arrowlength=0.6,arrowinset=0]{->}(82,75)(82,67.5)
\psellipse(70,75)(7,3.5)
\rput(70,75){End pt1c}
\put(75,70.5){PAc}

\put(18,74.5){
\tiny{
\put(-12,2.3){\psframe[shadow=true,fillstyle=solid,fillcolor=vert,linewidth=.5pt](2,2)(16,5)
\put(2.5,3){8`bob1c@0.0}}
\pscircle[fillstyle=solid,fillcolor=green,linewidth=.5pt](1,3.5){1.2}
\rput(1,3.5){8}
}}

%%------------ligne fabrication : Decoupe2

\put(0,-20){\put(-12,0){
\put(15.5,74.5){$m$}
\psellipse(35,75)(7,3.5)
\rput(35,75){Bobbin 2c}
\psline[linewidth=.7pt,arrowsize=0.12cm,arrowlength=0.6,arrowinset=0,linestyle=dashed]{->}(20,75)(28,75)
\put(40,70.5){PAc}
\psline[linewidth=.7pt,arrowsize=0.12cm,arrowlength=0.6,arrowinset=0]{->}(42,75)(55.25,75)
\put(53,20){
\psframe[linewidth=.7pt](2.25,52.5)(15.75,57.5)
\psframe[linewidth=.7pt](3,53.25)(15,56.75)
\rput(9,55){cutting 2}
\psframe[fillstyle=solid,fillcolor=bleu,linewidth=.2pt](4,51.5)(14,53.5)
\rput(9,52.5){\tiny {cutting 2}}
}
}
\psline[linewidth=.7pt,arrowsize=0.12cm,arrowlength=0.6,arrowinset=0]{->}(56.6,75)(63,75)
\psline[linewidth=.7pt,arrowsize=0.12cm,arrowlength=0.6,arrowinset=0]{-}(77,75)(82,75)
\psline[linewidth=.7pt,arrowsize=0.12cm,arrowlength=0.6,arrowinset=0]{->}(82,75)(82,81.5)
\psellipse(70,75)(7,3.5)
\rput(70,75){End pt2c}
\put(75,70.5){PAc}

\put(18,74.5){
\tiny{
\put(-12,2.3){\psframe[shadow=true,fillstyle=solid,fillcolor=vert,linewidth=.5pt](2,2)(16,5)
\put(2.5,3){5`bob2c@0.0}}
\pscircle[fillstyle=solid,fillcolor=green,linewidth=.5pt](1,3.5){1.2}
\rput(1,3.5){8}
}}
}

%%--------------ligne DB + couture

%%%%%--DB
\put(30.5,24){
\rput(19.5,41){\psovalbox{DB}}
\put(23,37.5){DB}
\psline[linewidth=.7pt,arrowsize=0.12cm,arrowlength=0.6,arrowinset=0]{<->}(19.5,38.5)(19.5,33.5)
\psline[linewidth=.7pt,arrowsize=0.12cm,arrowlength=0.6,arrowinset=0]{<->}(19.5,43.5)(19.5,47.5)
\psline[linewidth=.7pt,arrowsize=0.12cm,arrowlength=0.6,arrowinset=0]{<->}(23.6,41)(44.8,41)
%%--token
\put(14,37.5){\tiny{
\put(-20,0){
\psframe[shadow=true,fillstyle=solid,fillcolor=vert,linewidth=.5pt](2,0)(20,7)
\put(2.5,5){1`DB1@0.0+++}
\put(2.5,3){1`DB2@0.0+++}
\put(2.5,1){1`DB3@0.0}
}
\pscircle[fillstyle=solid,fillcolor=green,linewidth=.5pt](1,3.5){1.2}
\rput(1,3.5){3}
}}

\put(42.5,-14){
\psframe[linewidth=.7pt](2.25,52.5)(15.75,57.5)
\psframe[linewidth=.7pt](3,53.25)(15,56.75)
\rput(9,55){sewing}
\psframe[fillstyle=solid,fillcolor=bleu,linewidth=.2pt](4,51.5)(14,53.5)
\rput(9,52.5){\tiny {sewing}}
\psline[linewidth=.7pt,arrowsize=0.12cm,arrowlength=0.6,arrowinset=0]{->}(15.75,55)(24.2,55)
}

%%%%%--sewing
\put(55,0){
\psellipse(19.5,41)(8,3.5)
\rput(19.5,41){\scriptsize headdressc}
\put(25.8,37){PAc}
\psline[linewidth=.7pt,arrowsize=0.12cm,arrowlength=0.6,arrowinset=0,linestyle=dashed]{->}(28,41)(35.8,41)
}
}

}}
}
}
\end{pspicture}
}
\caption{Global view of the supply chain process}\label{Fig:hier_vue1}
\end{figure}
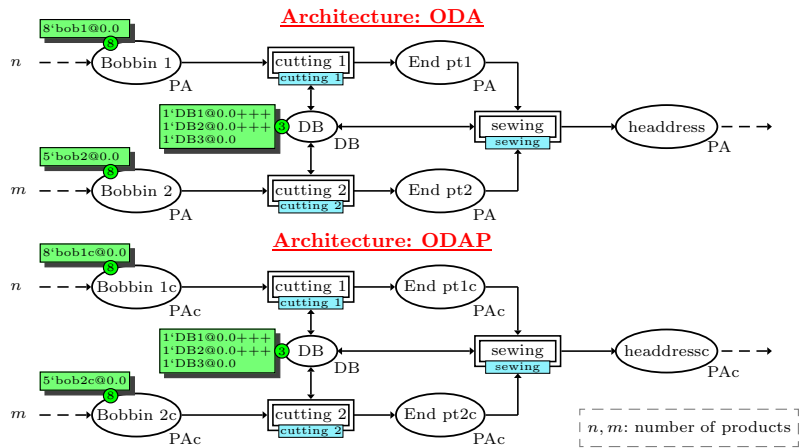

The architecture described in the previous paragraph corresponds to the Optimal Distributed Architecture (ODA), without any possibility to allocate data fragments\footnote{Products are only given an identifier (stored in a RFID tag) referring to a database.} on communicating products. In the second stage, we allow data fragment allocation on the products (according to their memory capacity and characteristics). We denote this architecture: ODAP (Optimal Distributed Architecture considering communicating Products). The idea is to highlight the potential benefits that could be achieved regarding a supply chain via such an ODAP architecture, where manufactured products act as mobile databases as opposed to a classic architecture (ODA). The figure~\ref{Fig:hier_vue1} illustrates in form of Petri Nets the synoptic of the part of the supply chain peculiar to our application. It consists of the ODA architecture implementing classic product and of the ODAP architecture implementing communicating product. %The purpose is to compare both architectures (ODA-ODAP) for a same logistic process: same inter-arrival law of products (see $\lambda 1$ and $\lambda 2$ on the figure~\ref{Fig:hier_vue1}), same operation duration\ldots
Let us note that we design the Petri Net model by working on hierarchical views. Consequently, the distribution aspect will be detailed in the lower views (in the section~\ref{Sec4:EvalSyst:CPN}).

\vspace{-0.5cm}
%%%%%%---- Vue 1_Hierarchy------%%%%%%%

\section{ODA and ODAP architecture modeling}\label{Sec4:EvalSyst}

\subsection{Architecture}\label{Sec4:EvalSyst:Eval_Arch}
\begin{figure}[ht]
\centering
\scalebox{1}{
\psset{xunit=1mm,yunit=1mm,runit=1mm}
\begin{pspicture}(0,18.5)(133,60)
%\psframe(0,18.5)(133,60)

\put(-5,0){
%----Physical Architecture
\psframe[framearc=.3,linecolor=gray,linestyle=dashed](5,18.5)(51,60)
\put(-13,0){\psframe[framearc=.3,linecolor=gray,linestyle=dashed](71,18.5)(117,60)}

\put(8,20){\scriptsize \emph{\textcolor{gray}{OPNET simulator}}}
\put(61,22.5){\scriptsize \emph{\textcolor{gray}{CPN Tools}}}
\put(61,20){\scriptsize \emph{\textcolor{gray}{simulator}}}

\put(0,2){
\rput(0,14){
\psline[linewidth=.5pt](14.5,39)(14.5,32)
\psline[linewidth=.5pt](28,39)(28,32)
\psline[linewidth=.5pt](41.5,39)(41.5,32)
\psline[linewidth=.5pt](14.5,39)(41.5,39)
\rput(21.25,40){\tiny \textcolor{gray}{\emph{10M}}}
\rput(34.75,40){\tiny \textcolor{gray}{\emph{10M}}}
\put(2,0){
\rput(15.5,35.5){\tiny \textcolor{gray}{\emph{10M}}}
\rput(29,35.5){\tiny \textcolor{gray}{\emph{10M}}}
\rput(42.5,35.5){\tiny \textcolor{gray}{\emph{10M}}}
}
\rput(14.5,39){\includegraphics[height=.3cm]{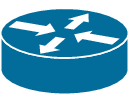}}
\rput(28,39){\includegraphics[height=.3cm]{router.eps}}
\rput(41.5,39){\includegraphics[height=.3cm]{router.eps}}
}

\psline[linewidth=.5pt](11.5,39)(14.5,46)
\psline[linewidth=.5pt](17.5,39)(14.5,46)
\rput(11.5,39){\includegraphics[height=.3cm]{pc.eps}}
\rput(11.5,35.5){\tiny Mach1}
\rput(17.5,39){\includegraphics[height=.3cm]{file_server.eps}}
\rput(17.5,35.5){\tiny DB1}
\psline[linewidth=.5pt](25,39)(28,46)
\psline[linewidth=.5pt](31,39)(28,46)
\rput(25,39){\includegraphics[height=.3cm]{pc.eps}}
\rput(25,35.5){\tiny Mach2}
\rput(31,39){\includegraphics[height=.3cm]{file_server.eps}}
\rput(31,35.5){\tiny DB2}
\psline[linewidth=.5pt](38.5,39)(41.5,46)
\psline[linewidth=.5pt](44.5,39)(41.5,46)
\rput(38.5,39){\includegraphics[height=.3cm]{pc.eps}}
\rput(38.5,35.5){\tiny Mach3}
\rput(44.5,39){\includegraphics[height=.3cm]{file_server.eps}}
\rput(44.5,35.5){\tiny DB3}

\rput(0,7){
\rput(14.5,39){\includegraphics[height=.3cm]{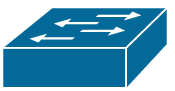}}
\rput(28,39){\includegraphics[height=.3cm]{switch.eps}}
\rput(41.5,39){\includegraphics[height=.3cm]{switch.eps}}
\rput(10.5,35.5){\tiny \textcolor{gray}{\emph{10M}}}
\rput(18.5,35.5){\tiny \textcolor{gray}{\emph{10M}}}
\put(13.5,0){
\rput(10.5,35.5){\tiny \textcolor{gray}{\emph{10M}}}
\rput(18.5,35.5){\tiny \textcolor{gray}{\emph{10M}}}
}
\put(27,0){
\rput(10.5,35.5){\tiny \textcolor{gray}{\emph{10M}}}
\rput(18.5,35.5){\tiny \textcolor{gray}{\emph{10M}}}
}
}
}
\psframe(7,35)(49,58)
\psline[linewidth=.7pt,arrowsize=0.18cm,arrowlength=0.8,arrowinset=0]{->}(28,35)(28,30)

%\rput(16.5,56){\tiny Definition of the}
%\rput(16.5,53){\tiny ODA architecture}
%\psframe(30,50)(49,58)
%\rput(39.5,56){\tiny Transaction}
%\rput(39.5,53){\tiny specification}
\put(0,-3){
\put(0,-15){
\pspolygon(28,42)(36,45)(28,48)(20,45)(28,42)
\rput(28,45){\emph{\tiny Simulations}}
}
%\psline(39.5,45)(39.5,53)
%\put(40.5,50.33){\tiny{\textcolor{gray}{query}}}
%\put(40.5,47.66){\tiny{\textcolor{gray}{patterns}}}
%\psline[linewidth=.7pt,arrowsize=0.18cm,arrowlength=0.8,arrowinset=0]{->}(39.5,45)(36,45)
%\psline[linewidth=.7pt](16.5,45)(16.5,53)
%\put(7,50.33){\tiny{\textcolor{gray}{physical}}}
%\put(7,47.66){\tiny{\textcolor{gray}{architec.}}}
%\psline[linewidth=.7pt,arrowsize=0.18cm,arrowlength=0.8,arrowinset=0]{->}(16.5,45)(20,45)
%\psframe(17.5,37)(38.5,29.5)
%\psline[linewidth=.7pt,arrowsize=0.18cm,arrowlength=0.8,arrowinset=0]{->}(28,42)(28,37)
%\rput(28,35){\tiny Results from a}
%\rput(28,32){\tiny set of simulations}
%\psline[linewidth=.7pt](28,29.5)(28,26)
\put(-1.5,0){
\psline[linewidth=.7pt](37.5,30)(56,30)
\psline[linewidth=.7pt](56,30)(56,45)
\put(54.10,31){\rotatebox{90}{\tiny{\textcolor{red}{\emph{Round trip time}}}}}
\put(56.75,33){\rotatebox{90}{\tiny{\textcolor{red}{statistics}}}}
}
}

\psline[linewidth=.7pt,arrowsize=0.18cm,arrowlength=0.8,arrowinset=0]{->}(54.5,42)(73,42)

\put(53,0){
\psframe(7,50)(26,58)
\rput(16.5,56){\tiny Definition of the}
\rput(16.5,53){\tiny Supply chain}
\psframe(30,50)(49,58)
\put(0,-.5){
\rput(39.5,57){\tiny ODA and ODAP}
\rput(39.5,54.5){\tiny distribution}
\rput(39.5,52){\tiny specification}
}
\put(0,-3){
\pspolygon(28,42)(36,45)(28,48)(20,45)(28,42)
\rput(28,45){\emph{\tiny Simulations}}
\psline[linewidth=.7pt](39.5,45)(39.5,53)
\put(40.5,50.33){\tiny{\textcolor{gray}{possible}}}
\put(40.5,47.66){\tiny{\textcolor{gray}{combinat.}}}
\psline[linewidth=.7pt,arrowsize=0.18cm,arrowlength=0.8,arrowinset=0]{->}(39.5,45)(36,45)
\psline(16.5,50.5)(16.5,53)
\put(9,50.33){\tiny{\textcolor{gray}{tasks}}}
\put(9,47.66){\tiny{\textcolor{gray}{definition}}}
\psline[linewidth=.7pt](16.5,50.5)(28,50.5)
\psline[linewidth=.7pt,arrowsize=0.18cm,arrowlength=0.8,arrowinset=0]{->}(28,50.5)(28,48)
\psframe(17.5,37)(38.5,29.5)
\psline[linewidth=.7pt,arrowsize=0.18cm,arrowlength=0.8,arrowinset=0]{->}(28,42)(28,37)
\rput(28,35){\tiny Results from a}
\rput(28,32){\tiny set of simulations}
\psline[linewidth=.7pt,arrowsize=0.18cm,arrowlength=0.8,arrowinset=0]{->}(38.5,33.5)(55,33.5)
\put(10,5){
\rput(36.75,26.125){\tiny{\textcolor{gray}{Total time}}}
\rput(36.75,24.125){\tiny{\textcolor{gray}{to produce}}}
\rput(36.75,22.125){\tiny{\textcolor{gray}{$j$ headdresses}}}
}

\put(39.5,12.5){
\psframe(15.5,26.25)(35.5,15.75)
\psframe(16.25,25.5)(34.75,16.5)
\rput(25.5,23.5){\tiny Comparison}
\rput(25.5,21){\tiny between}
\rput(25.5,18.5){\tiny ODA and ODAP}
}

}

}

}
\end{pspicture}
}
\caption{Usage process of evaluation architecture}\label{Fig:Syst_Eval}
\end{figure}

A description of how the assessment and the comparison are undertaken of both architectures (ODA and ODAP) is proposed in this section. The evaluation architecture relies on two discrete events simulators and its usage process is depicted on the figure~\ref{Fig:Syst_Eval}. This architecture is composed by two sub-systems. The first one is a tool for editing, simulating, and analyzing Colored Petri Nets (CPN tools). As shown on the figure~\ref{Fig:Syst_Eval}, the supply chain behavior is simulated via this tool, or at least the considered part which has been exposed in the section~\ref{Sec3:StudyCase:adaptation} (operation $1$, $2$ and $3$). It allows to deal with sharing of physical resources taking place into the system (databases, machines, manufactured products\ldots), operation times, queuing tasks, times for reading/writing information on databases or still on manufactured products (considering the ODAP architecture) and so on. The ODA and ODAP distribution patterns are specified in this tool. Let us note that for the ODAP architecture, all the possible combination of distribution between the product and the distributed fixed databases are realized (i.e $2^{k}$ possibilities with $k$ the total number of fragments). The second tool is the OPNET network simulator which is primarily aimed at developing and validating network protocols. However, it allows estimating various parameters on specific case studies, such as network times, overload traffic, equipment processing times, battery life, etc. In our study, the OPNET tool is used for assessing the \emph{round trip time}\footnote{Time between sending the first packet of the request and receiving the last packet of the response} to achieve read/write queries on databases, considering the ODA architecture (we do not take into account the possibility to read and write the product). To do this, the physical architecture and the distribution adopted in the section~\ref{Sec3:StudyCase} have to be specified in OPNET. The resulting times are then injected into CPN Tools. The next sections introduce the methodology employed for assessing the ODA and ODAP architectures.

\subsection{Estimated "round trip times" of ODA via OPNET}\label{Sec4:EvalSyst:OPNET}
First, the network interconnecting the client machines and the fixed databases is defined in OPNET (see figure~\ref{Fig:Syst_Eval}). Thereafter, it is necessary to create system partitions on each server in order to allocate the data fragments (i.e the ODA distribution specified in the figure~\ref{Fig:arch_phys}). Therefore, a replicating protocol has to be implemented owing to the replication of $F5$, $F7$ and $F8$. In our application, we implement Synchronous and Primary copy mechanisms described in the section~\ref{Sec2:DDBS}. Subsequently, it is necessary to specify the applicative exchanges between equipments, i.e the query pattern (read/write) performed by each client machine on the databases. To do this, three models are implemented in OPNET: \emph{Task}, \emph{Application} and \emph{Profile} models. %Both above points are the needed inputs for simulating the ODA architecture (see figure~\ref{Fig:Syst_Eval}).
Finally, it is possible to estimate the \emph{round trip time} for a specific query sent from a client to a database.

Statistical tools are available in OPNET for computing averages, variances or still confidence intervals based on a set of simulations. In our study, $50$ simulations have been running for a same scenario and then, both the \emph{round trip time} average and the statistical variance have been extracted for each query. For instant, the table~\ref{Table:tps_Opnet} gives the \emph{round trip time} induced by a read (R) or write (W) query on $F1$ (fragment allocated to DB1).
The machine $1$ requires $3.6ms$ on average with a variance of $9 \mu s$ to access to this fragment and spends $7.6ms$ and $7 \mu s$ respectively to write it. %The average and variance statistical times related to this query are about  and  respectively. %For the replicated fragments, it is possible to read any databases carrying replica (e.g the machine $1$ can reach either DB1 or DB3 for reading F8). Nonetheless, the only possibility for writing fragments is via the Primary copy fragment. The way for injecting times into Petri Nets is describes in the following section.

\vspace{-.15cm}
\begin{table}[htb]
  \caption{Evaluated times regarding access query patterns: S-Ue}
  \label{Table:tps_Opnet}
  \begin{center}
    \begin{tabular}{cc ccc ccc ccc}
    \cline{3-11}
          &  & \multicolumn{3}{c}{Machine 1} & \multicolumn{3}{c}{Machine 2} & \multicolumn{3}{c}{Machine 3}\\
          &  & DB1 & DB2 & DB3 & DB1 & DB2 & DB3 & DB1 & DB2 & DB3 \\
          \cline{1-11}

          \multirow{2}{*}{\rotatebox{90}{F1}} & R & \scriptsize{$3.6ms,0.009\mu s$} & $\times$ & $\times$ & $\times$ & $\times$ & $\times$ & \scriptsize{$7.8ms,0.012\mu s$} & $\times$ & $\times$\\
                & W & \scriptsize{$7.6ms,0.007\mu s$} & $\times$ & $\times$& $\times$ & $\times$ & $\times$ & \scriptsize{$11.3ms,0.015\mu s$} & $\times$ & $\times$ \\
%
%          \multirow{2}{*}{\rotatebox{90}{F2}} & R & & & & & & \scriptsize{$8.6,0.032$} & & & \scriptsize{$5.15,0.010$} \\
%                & U & & & & & & \scriptsize{$11.8,0.031$} & & & \scriptsize{$9.4,0.016$} \\
%
%                \multicolumn{11}{c}{\vdots} \\

%          \multirow{2}{*}{\rotatebox{90}{F8}} & R & \scriptsize{$3.9,0.009$} & \scriptsize{$9.7,0.015$} & & \scriptsize{$7.6,0.010$} & \scriptsize{$5.13,0.013$} & & & & \\
%                & U & & \scriptsize{$15.8,0.029$} & & & \scriptsize{$11.6,0.013$} & & & \scriptsize{$15.5,0.023$} & \\
         \cline{1-11}
    \end{tabular}
  \end{center}
\end{table}

\vspace{-.75cm}

\subsection{Petri Nets: ODA and ODAP architectures}\label{Sec4:EvalSyst:CPN}
\begin{figure}[ht]
\centering
\scalebox{.9}{
\psset{xunit=1mm,yunit=1mm,runit=1mm}
\begin{pspicture}(10,-14.5)(122,85)
%\psframe(10,-14.5)(122,85)

\definecolor{vert}{rgb}{0.46,1.00,0.46}
\definecolor{bleu}{rgb}{0.56,0.95,0.99}
\definecolor{bleuc}{rgb}{0.84,0.98,0.99}
\definecolor{jauneF}{rgb}{0.98,0.78,0.11}

\put(-43,0){
\put(-10,0){
\tiny{
\psellipse[linecolor=blue](90,82)(6,3)
\rput(90,82){\textcolor{blue}{Bobbin 1}}
\psline[linewidth=.7pt,arrowsize=0.12cm,arrowlength=0.6,arrowinset=0]{->}(90,79)(90,75)
\put(90.25,77){\tiny \emph{pa}}
\put(95,78){\textcolor{blue}{PA}}
\put(17,0){
\psframe[linewidth=.2,linecolor=blue,fillstyle=solid,fillcolor=bleuc](78.3,81)(81.7,83)
\rput(80,82){\tiny \textcolor{blue}{In}}
}

\rput(90,73){\psframebox{start Rd}}
\psline[linewidth=.7pt,arrowsize=0.12cm,arrowlength=0.6,arrowinset=0]{->}(90,70.95)(90,65)
\put(90.25,67.95){\tiny \emph{bob1}}
\rput(79.2,71){\tiny \textcolor{red}{\emph{fd1R\_DB()}}}

%\psellipse(61,68)(7,3.5)
%\rput(61,69){cutting}
%\rput(61,67){machine}
%\rput(72,76){\tiny \emph{l}}
%\put(65,63){M}

\rput(90,63){\psovalbox{b1 read}}
\psline[linewidth=.7pt,arrowsize=0.12cm,arrowlength=0.6,arrowinset=0]{->}(90,60.75)(90,54.75)
\put(95,59){PA}
\put(90.25,58){\tiny \emph{pa}}

\put(-22,10){
\rput(90,53){\psdblframebox{read DB}}
\psline[linewidth=.7pt,arrowsize=0.12cm,arrowlength=0.6,arrowinset=0]{->}(90,50.2)(90,46.5)
\psline[linewidth=.7pt,arrowsize=0.12cm,arrowlength=0.6,arrowinset=0]{<-}(90,55.5)(90,59.6)
\psframe[fillstyle=solid,fillcolor=bleu,linewidth=.2pt](85,49.5)(95,51.5)
\rput(90,50.5){\tiny {read DB}}
}

\psellipse(68,53)(7,3.5)
\rput(68,54){Info DB}
\rput(68,52){aft. read}
\psline[linewidth=.7pt,arrowsize=0.12cm,arrowlength=0.6,arrowinset=0]{->}(75,53)(85,53)
%\psline[linewidth=.7pt,arrowsize=0.12cm,arrowlength=0.6,arrowinset=0]{<-}(95,53)(105,53)
\put(71.5,56.5){FRAG}

\put(0,20){
\psellipse(68,53)(7,3.5)
\rput(68,54){Info DB}
\rput(68,52){bef. read}
\psline[linewidth=.7pt,arrowsize=0.12cm,arrowlength=0.6,arrowinset=0]{<-}(75,53)(84,53)
%\psline[linewidth=.7pt,arrowsize=0.12cm,arrowlength=0.6,arrowinset=0]{->}(95.8,53)(105,53)
\put(71.5,56.5){FRAG}
}

\rput(90,53){\psframebox{end Rd}}
\psline[linewidth=.7pt,arrowsize=0.12cm,arrowlength=0.6,arrowinset=0]{->}(90,51)(90,46.5)
%\psframe[fillstyle=solid,fillcolor=bleu,linewidth=.2pt](85,49.5)(95,51.5)
%\rput(90,50.5){\tiny {read DB}}
\put(90.25,48.75){\tiny \emph{bob1}}
\rput(79.2,51){\tiny \emph{\textcolor{red}{fd1R\_DB()}}}

\psellipse(90,43)(8,3.5)
\rput(90,44){AP + info}
\rput(90,42){operation}
\psline[linewidth=.7pt,arrowsize=0.12cm,arrowlength=0.6,arrowinset=0]{->}(90,39.75)(90,34.75)
\put(90.25,37){\tiny \emph{pa}}
%\put(55,47.5){\tiny \emph{m}}
\put(96.5,38.5){PA}

\rput(90,33){\psframebox[linecolor=gray]{cut}}
\psline[linewidth=.7pt,arrowsize=0.12cm,arrowlength=0.6,arrowinset=0]{->}(90,31)(90,26.25)
\put(90.25,28.675){\tiny \emph{19`pt1}}
\put(-15,-5){
\put(97,36){\tiny \emph{\textcolor{gray}{@++}}}
\put(91,34){\tiny \emph{\textcolor{gray}{operTime(pa)}}}
}

\put(-5,0){
\rput(73,33){\psovalbox{Ma.1}}
%\psline[linewidth=.7pt,arrowsize=0.12cm,arrowlength=0.6,arrowinset=0]{->}(90,60.75)(90,54.75)
\put(77,30){M}
\psline[linewidth=.7pt,arrowsize=0.12cm,arrowlength=0.6,arrowinset=0]{<->}(78,33)(92,33)
\rput(84,34){\tiny \emph{m1}}
}

\tiny{
\put(59,38){
\rput(-3.5,3.5){\psovalbox[linecolor=blue]{\textcolor{blue}{DB}}}
\put(-2,-1){\textcolor{blue}{DB}}
\psline[linewidth=.7pt,arrowsize=0.12cm,arrowlength=0.6,arrowinset=0]{<-}(-3.5,7.4)(-3.5,25)
\psline[linewidth=.7pt,arrowsize=0.12cm,arrowlength=0.6,arrowinset=0]{->}(-3.5,25)(2.75,25)
\psline[linewidth=.7pt,arrowsize=0.12cm,arrowlength=0.6,arrowinset=0]{<-}(-3.5,1.5)(-3.5,-32)
\psline[linewidth=.7pt,arrowsize=0.12cm,arrowlength=0.6,arrowinset=0]{->}(-3.5,-32)(2.3,-32)
\put(-2,5.2){
\psframe[linewidth=.2,linecolor=blue,fillstyle=solid,fillcolor=bleuc](-3.7,-0.2)(0.7,2.2)
\rput(-1.5,.75){\textcolor{blue}{\tiny I/O}}}
\tiny{
\psframe[shadow=true,fillstyle=solid,fillcolor=vert,linewidth=.5pt](2,0)(20,7)
\put(2.5,5){1`DB1@0.0+++}
\put(2.5,3){1`DB2@0.0+++}
\put(2.5,1){1`DB3@0.0}
\pscircle[fillstyle=solid,fillcolor=green,linewidth=.5pt](1,3.5){1.2}
\rput(1,3.5){3}
}}
}

\rput(90.25,24){\psovalbox{pt1}}
%\psellipse(90,23)(7,3.5)
%\rput(90,24){pieces 2}
%\rput(90,22){textile 2}
\psline[linewidth=.7pt,arrowsize=0.12cm,arrowlength=0.6,arrowinset=0]{->}(90,21.6)(90,17.85)
\put(94,20.5){PA}
\put(90.25,19.725){\tiny \emph{pa}}

\put(0,-57){
\rput(90,73){\psframebox{start Wr}}
\psline[linewidth=.7pt,arrowsize=0.12cm,arrowlength=0.6,arrowinset=0]{->}(90,70.95)(90,65.8)
\put(90.25,67.95){\tiny \emph{pt1}}
\rput(79.2,71){\tiny \emph{\textcolor{red}{fd1W\_DB()}}}

%\psellipse(61,68)(7,3.5)
%\rput(61,69){cutting}
%\rput(61,67){machine}
%\rput(72,76){\tiny \emph{l}}
%\put(65,63){M}

\psellipse(90,63)(6,3)
\rput(90,63){pt1 write}
%\rput(90,63){\psovalbox{pt1c write}}
\psline[linewidth=.7pt,arrowsize=0.12cm,arrowlength=0.6,arrowinset=0]{->}(90,60)(90,54.75)
\put(95.5,59){PA}
\put(90.25,57.95){\tiny \emph{pa}}

\put(-22,10){
\rput(90,53){\psdblframebox{write DB}}
\psline[linewidth=.7pt,arrowsize=0.12cm,arrowlength=0.6,arrowinset=0]{->}(90,50.2)(90,46.5)
\psline[linewidth=.7pt,arrowsize=0.12cm,arrowlength=0.6,arrowinset=0]{<-}(90,55.5)(90,59.6)
\psframe[fillstyle=solid,fillcolor=bleu,linewidth=.2pt](85,49.5)(95,51.5)
\rput(90,50.5){\tiny {write DB}}
}

\psellipse(68,53)(7,3.5)
\rput(68,54){Info DB}
\rput(68,52){aft. write}
\psline[linewidth=.7pt,arrowsize=0.12cm,arrowlength=0.6,arrowinset=0]{->}(75,53)(85,53)
%\psline[linewidth=.7pt,arrowsize=0.12cm,arrowlength=0.6,arrowinset=0]{<-}(95,53)(105,53)
\put(71.5,56.5){FRAG}

\put(0,20){
\psellipse(68,53)(7,3.5)
\rput(68,54){Info DB}
\rput(68,52){bef. write}
\psline[linewidth=.7pt,arrowsize=0.12cm,arrowlength=0.6,arrowinset=0]{<-}(75,53)(84,53)
%\psline[linewidth=.7pt,arrowsize=0.12cm,arrowlength=0.6,arrowinset=0]{->}(95.8,53)(105,53)
\put(71.5,56.5){FRAG}
}

%\put(22,10){
%\rput(90,53){\psdblframebox{write AP}}
%\psline[linewidth=.7pt,arrowsize=0.12cm,arrowlength=0.6,arrowinset=0]{->}(90,50.2)(90,46.5)
%\psline[linewidth=.7pt,arrowsize=0.12cm,arrowlength=0.6,arrowinset=0]{<-}(90,55.5)(90,59.6)
%\psframe[fillstyle=solid,fillcolor=bleu,linewidth=.2pt](85,49.5)(95,51.5)
%\rput(90,50.5){\tiny {write AP}}
%}

%\put(44,0){
%\psellipse(68,53)(7,3.5)
%\rput(68,54){Info AP}
%\rput(68,52){aft. write}
%\put(71.5,56.5){FRAG}
%
%\put(0,20){
%\psellipse(68,53)(7,3.5)
%\rput(68,54){Info AP}
%\rput(68,52){bef. write}
%%\psline[linewidth=.7pt,arrowsize=0.12cm,arrowlength=0.6,arrowinset=0]{<->}(75,53)(83,53)
%%\psline[linewidth=.7pt,arrowsize=0.12cm,arrowlength=0.6,arrowinset=0]{<->}(97,53)(103.5,53)
%\put(71.5,56.5){FRAG}
%}
%}

\rput(90,53){\psframebox{end Wr}}
\psline[linewidth=.7pt,arrowsize=0.12cm,arrowlength=0.6,arrowinset=0]{->}(90,51)(90,47.4)
%\psframe[fillstyle=solid,fillcolor=bleu,linewidth=.2pt](85,49.5)(95,51.5)
%\rput(90,50.5){\tiny {read DB}}
\put(90.25,49.2){\tiny \emph{pt1}}
\rput(79.2,51){\tiny \emph{\textcolor{red}{fd1W\_DB()}}}
}

\put(0,-15){
\rput(90,3){\psovalbox[linecolor=blue]{\textcolor{blue}{End pt1}}}
\put(77,2){\textcolor{blue}{LOT}}
\rput(63,5){\textcolor{red}{\normalsize{\textbf{\emph{ODA}}}}}
\rput(63,2){\textcolor{red}{\normalsize{\textbf{\emph{architecture}}}}}
\tiny
\put(17.5,0){
\psframe[linewidth=.2,linecolor=blue,fillstyle=solid,fillcolor=bleuc](78.4,2)(82.6,4)
\rput(80.5,3){\tiny \textcolor{blue}{Out}}}
}
}
}

\put(73,79){
\tiny{
\put(-16,0){\psframe[shadow=true,fillstyle=solid,fillcolor=vert,linewidth=.5pt](2,2)(16,5)
\put(2.5,3){8`bob1@0.0}}
\pscircle[fillstyle=solid,fillcolor=green,linewidth=.5pt](1,3.5){1.2}
\rput(1,3.5){8}
}}

\put(57,27){
\tiny{
\put(-12,-2.5){\psframe[shadow=true,fillstyle=solid,fillcolor=vert,linewidth=.5pt](4.5,2)(16,5)
\put(5,3){1`m1@0.0}}
\pscircle[fillstyle=solid,fillcolor=green,linewidth=.5pt](1,3.5){1.2}
\rput(1,3.5){1}
}}

}

\psline[linewidth=1.5pt,linecolor=red,linestyle=dashed](50,-14.5)(50,85)

\put(10,0){
\put(-10,0){
\tiny{
\psellipse[linecolor=blue](90,82)(6,3)
\rput(90,82){\textcolor{blue}{Bobbin 1c}}
\psline[linewidth=.7pt,arrowsize=0.12cm,arrowlength=0.6,arrowinset=0]{->}(90,79)(90,75)
\put(90.25,77){\tiny \emph{pac}}
\put(95,78){\textcolor{blue}{PAc}}
\put(17,0){
\psframe[linewidth=.2,linecolor=blue,fillstyle=solid,fillcolor=bleuc](78.3,81)(81.7,83)
\rput(80,82){\tiny \textcolor{blue}{In}}
}

\rput(90,73){\psframebox{start Rd}}
\psline[linewidth=.7pt,arrowsize=0.12cm,arrowlength=0.6,arrowinset=0]{->}(90,70.95)(90,65)
\put(90.25,67.95){\tiny \emph{bob1c}}
\rput(79.2,71){\tiny \emph{\textcolor{red}{fd1R\_DB()}}}
\rput(100.8,71){\tiny \emph{\textcolor{red}{fd1R\_PA()}}}

%\psellipse(61,68)(7,3.5)
%\rput(61,69){cutting}
%\rput(61,67){machine}
%\rput(72,76){\tiny \emph{l}}
%\put(65,63){M}

\rput(90,63){\psovalbox{b1c read}}
\psline[linewidth=.7pt,arrowsize=0.12cm,arrowlength=0.6,arrowinset=0]{->}(90,60.75)(90,54.75)
\put(95,59){PAc}
\put(90.25,58){\tiny \emph{pac}}

\put(-22,10){
\rput(90,53){\psdblframebox{read DB}}
\psline[linewidth=.7pt,arrowsize=0.12cm,arrowlength=0.6,arrowinset=0]{->}(90,50.2)(90,46.5)
\psline[linewidth=.7pt,arrowsize=0.12cm,arrowlength=0.6,arrowinset=0]{<-}(90,55.5)(90,59.6)
\psframe[fillstyle=solid,fillcolor=bleu,linewidth=.2pt](85,49.5)(95,51.5)
\rput(90,50.5){\tiny {read DB}}
}

\psellipse(68,53)(7,3.5)
\rput(68,54){Info DB}
\rput(68,52){aft. read}
\psline[linewidth=.7pt,arrowsize=0.12cm,arrowlength=0.6,arrowinset=0]{->}(75,53)(85,53)
\psline[linewidth=.7pt,arrowsize=0.12cm,arrowlength=0.6,arrowinset=0]{<-}(95,53)(105,53)
\put(71.5,56.5){FRAG}

\put(0,20){
\psellipse(68,53)(7,3.5)
\rput(68,54){Info DB}
\rput(68,52){bef. read}
\psline[linewidth=.7pt,arrowsize=0.12cm,arrowlength=0.6,arrowinset=0]{<-}(75,53)(84,53)
\psline[linewidth=.7pt,arrowsize=0.12cm,arrowlength=0.6,arrowinset=0]{->}(95.8,53)(105,53)
\put(71.5,56.5){FRAG}
}

\put(22,10){
\rput(90,53){\psdblframebox{read AP}}
\psline[linewidth=.7pt,arrowsize=0.12cm,arrowlength=0.6,arrowinset=0]{->}(90,50.2)(90,46.5)
\psline[linewidth=.7pt,arrowsize=0.12cm,arrowlength=0.6,arrowinset=0]{<-}(90,55.5)(90,59.6)
\psframe[fillstyle=solid,fillcolor=bleu,linewidth=.2pt](85,49.5)(95,51.5)
\rput(90,50.5){\tiny {read AP}}
}

\put(44,0){
\psellipse(68,53)(7,3.5)
\rput(68,54){Info AP}
\rput(68,52){aft. read}
\put(71.5,56.5){FRAG}

\put(0,20){
\psellipse(68,53)(7,3.5)
\rput(68,54){Info AP}
\rput(68,52){bef. read}
%\psline[linewidth=.7pt,arrowsize=0.12cm,arrowlength=0.6,arrowinset=0]{<->}(75,53)(83,53)
%\psline[linewidth=.7pt,arrowsize=0.12cm,arrowlength=0.6,arrowinset=0]{<->}(97,53)(103.5,53)
\put(71.5,56.5){FRAG}
}
}

\rput(90,53){\psframebox{end Rd}}
\psline[linewidth=.7pt,arrowsize=0.12cm,arrowlength=0.6,arrowinset=0]{->}(90,51)(90,46.5)
%\psframe[fillstyle=solid,fillcolor=bleu,linewidth=.2pt](85,49.5)(95,51.5)
%\rput(90,50.5){\tiny {read DB}}
\put(90.25,48.75){\tiny \emph{bob1c}}
\rput(79.2,51){\tiny \emph{\textcolor{red}{fd1R\_DB()}}}
\rput(100.8,51){\tiny \emph{\textcolor{red}{fd1R\_PA()}}}

\psellipse(90,43)(8,3.5)
\rput(90,44){PAc + info}
\rput(90,42){operation}
\psline[linewidth=.7pt,arrowsize=0.12cm,arrowlength=0.6,arrowinset=0]{->}(90,39.75)(90,34.75)
\put(90.25,37){\tiny \emph{pa}}
%\put(55,47.5){\tiny \emph{m}}
\put(96.5,38.5){PA}

\rput(90,33){\psframebox[linecolor=gray]{cut}}
\psline[linewidth=.7pt,arrowsize=0.12cm,arrowlength=0.6,arrowinset=0]{->}(90,31)(90,26.25)
\put(90.25,28.675){\tiny \emph{19`pt1}}
\put(-15,-5){
\put(97,36){\tiny \emph{\textcolor{gray}{@++}}}
\put(91,34){\tiny \emph{\textcolor{gray}{operTime(pac)}}}
}

\put(-5,0){
\rput(73,33){\psovalbox{Ma.1}}
%\psline[linewidth=.7pt,arrowsize=0.12cm,arrowlength=0.6,arrowinset=0]{->}(90,60.75)(90,54.75)
\put(77,30){M}
\psline[linewidth=.7pt,arrowsize=0.12cm,arrowlength=0.6,arrowinset=0]{<->}(78,33)(92,33)
\rput(81.5,34){\tiny \emph{m1}}
}

\tiny{
\put(59,38){
\rput(-3.5,3.5){\psovalbox[linecolor=blue]{\textcolor{blue}{DB}}}
\put(-2,-1){\textcolor{blue}{DB}}
\psline[linewidth=.7pt,arrowsize=0.12cm,arrowlength=0.6,arrowinset=0]{<-}(-3.5,7.4)(-3.5,25)
\psline[linewidth=.7pt,arrowsize=0.12cm,arrowlength=0.6,arrowinset=0]{->}(-3.5,25)(2.75,25)
\psline[linewidth=.7pt,arrowsize=0.12cm,arrowlength=0.6,arrowinset=0]{<-}(-3.5,1.5)(-3.5,-32)
\psline[linewidth=.7pt,arrowsize=0.12cm,arrowlength=0.6,arrowinset=0]{->}(-3.5,-32)(2.3,-32)
\put(-2,5.2){
\psframe[linewidth=.2,linecolor=blue,fillstyle=solid,fillcolor=bleuc](-3.7,-0.2)(0.7,2.2)
\rput(-1.5,.75){\textcolor{blue}{\tiny I/O}}}
\tiny{
\psframe[shadow=true,fillstyle=solid,fillcolor=vert,linewidth=.5pt](2,0)(20,7)
\put(2.5,5){1`DB1@0.0+++}
\put(2.5,3){1`DB2@0.0+++}
\put(2.5,1){1`DB3@0.0}
\pscircle[fillstyle=solid,fillcolor=green,linewidth=.5pt](1,3.5){1.2}
\rput(1,3.5){3}
}}
}

\rput(90.25,24){\psovalbox{pt1c}}
%\psellipse(90,23)(7,3.5)
%\rput(90,24){pieces 2}
%\rput(90,22){textile 2}
\psline[linewidth=.7pt,arrowsize=0.12cm,arrowlength=0.6,arrowinset=0]{->}(90,21.6)(90,17.85)
\put(94,20.5){PA}
\put(90.25,19.725){\tiny \emph{pac}}

\put(0,-57){
\rput(90,73){\psframebox{start Wr}}
\psline[linewidth=.7pt,arrowsize=0.12cm,arrowlength=0.6,arrowinset=0]{->}(90,70.95)(90,65.8)
\put(90.25,67.95){\tiny \emph{pt1c}}
\rput(79.2,71){\tiny \emph{\textcolor{red}{fd1W\_DB()}}}
\rput(100.8,71){\tiny \emph{\textcolor{red}{fd1W\_PA()}}}

%\psellipse(61,68)(7,3.5)
%\rput(61,69){cutting}
%\rput(61,67){machine}
%\rput(72,76){\tiny \emph{l}}
%\put(65,63){M}

\psellipse(90,63)(6,3)
\rput(90,63){pt1c write}
%\rput(90,63){\psovalbox{pt1c write}}
\psline[linewidth=.7pt,arrowsize=0.12cm,arrowlength=0.6,arrowinset=0]{->}(90,60)(90,54.75)
\put(95.5,59){PAc}
\put(90.25,57.95){\tiny \emph{pac}}

\put(-22,10){
\rput(90,53){\psdblframebox{write DB}}
\psline[linewidth=.7pt,arrowsize=0.12cm,arrowlength=0.6,arrowinset=0]{->}(90,50.2)(90,46.5)
\psline[linewidth=.7pt,arrowsize=0.12cm,arrowlength=0.6,arrowinset=0]{<-}(90,55.5)(90,59.6)
\psframe[fillstyle=solid,fillcolor=bleu,linewidth=.2pt](85,49.5)(95,51.5)
\rput(90,50.5){\tiny {write DB}}
}

\psellipse(68,53)(7,3.5)
\rput(68,54){Info DB}
\rput(68,52){aft. write}
\psline[linewidth=.7pt,arrowsize=0.12cm,arrowlength=0.6,arrowinset=0]{->}(75,53)(85,53)
\psline[linewidth=.7pt,arrowsize=0.12cm,arrowlength=0.6,arrowinset=0]{<-}(95,53)(105,53)
\put(71.5,56.5){FRAG}

\put(0,20){
\psellipse(68,53)(7,3.5)
\rput(68,54){Info DB}
\rput(68,52){bef. write}
\psline[linewidth=.7pt,arrowsize=0.12cm,arrowlength=0.6,arrowinset=0]{<-}(75,53)(84,53)
\psline[linewidth=.7pt,arrowsize=0.12cm,arrowlength=0.6,arrowinset=0]{->}(95.8,53)(105,53)
\put(71.5,56.5){FRAG}
}

\put(22,10){
\rput(90,53){\psdblframebox{write AP}}
\psline[linewidth=.7pt,arrowsize=0.12cm,arrowlength=0.6,arrowinset=0]{->}(90,50.2)(90,46.5)
\psline[linewidth=.7pt,arrowsize=0.12cm,arrowlength=0.6,arrowinset=0]{<-}(90,55.5)(90,59.6)
\psframe[fillstyle=solid,fillcolor=bleu,linewidth=.2pt](85,49.5)(95,51.5)
\rput(90,50.5){\tiny {write AP}}
}

\put(44,0){
\psellipse(68,53)(7,3.5)
\rput(68,54){Info AP}
\rput(68,52){aft. write}
\put(71.5,56.5){FRAG}

\put(0,20){
\psellipse(68,53)(7,3.5)
\rput(68,54){Info AP}
\rput(68,52){bef. write}
%\psline[linewidth=.7pt,arrowsize=0.12cm,arrowlength=0.6,arrowinset=0]{<->}(75,53)(83,53)
%\psline[linewidth=.7pt,arrowsize=0.12cm,arrowlength=0.6,arrowinset=0]{<->}(97,53)(103.5,53)
\put(71.5,56.5){FRAG}
}
}

\rput(90,53){\psframebox{end Wr}}
\psline[linewidth=.7pt,arrowsize=0.12cm,arrowlength=0.6,arrowinset=0]{->}(90,51)(90,47.4)
%\psframe[fillstyle=solid,fillcolor=bleu,linewidth=.2pt](85,49.5)(95,51.5)
%\rput(90,50.5){\tiny {read DB}}
\put(90.25,49.2){\tiny \emph{pt1c}}
\rput(79.2,51){\tiny \emph{\textcolor{red}{fd1W\_DB()}}}
\rput(100.8,51){\tiny \emph{\textcolor{red}{fd1W\_PA()}}}
}

\put(0,-15){
\rput(90,3){\psovalbox[linecolor=blue]{\textcolor{blue}{End pt1}}}
\put(77,2){\textcolor{blue}{LOT}}
\rput(63,5){\textcolor{red}{\normalsize{\textbf{\emph{ODAP}}}}}
\rput(63,2){\textcolor{red}{\normalsize{\textbf{\emph{architecture}}}}}
\tiny
\put(17.5,0){
\psframe[linewidth=.2,linecolor=blue,fillstyle=solid,fillcolor=bleuc](78.4,2)(82.6,4)
\rput(80.5,3){\tiny \textcolor{blue}{Out}}}
}
}
}

\put(72.5,79){
\tiny{
\put(-16,0){\psframe[shadow=true,fillstyle=solid,fillcolor=vert,linewidth=.5pt](2,2)(16,5)
\put(2.5,3){8`bob1c@0.0}}
\pscircle[fillstyle=solid,fillcolor=green,linewidth=.5pt](1,3.5){1.2}
\rput(1,3.5){8}
}}

\put(57,27){
\tiny{
\put(-12,-2.5){\psframe[shadow=true,fillstyle=solid,fillcolor=vert,linewidth=.5pt](4.5,2)(16,5)
\put(5,3){1`m1@0.0}}
\pscircle[fillstyle=solid,fillcolor=green,linewidth=.5pt](1,3.5){1.2}
\rput(1,3.5){1}
}}

}

\put(19,64){
\pspolygon[fillstyle=solid,fillcolor=yellow,linecolor=jauneF](3,0)(3.5,1.5)(6,2)(3.5,2.5)(3,4)(2.5,2.5)(0,2)(2.5,1.5)(3,0)
}

\put(19,7){
\pspolygon[fillstyle=solid,fillcolor=yellow,linecolor=jauneF](3,0)(3.5,1.5)(6,2)(3.5,2.5)(3,4)(2.5,2.5)(0,2)(2.5,1.5)(3,0)
}

\put(53,0){
\put(19,64){
\pspolygon[fillstyle=solid,fillcolor=yellow,linecolor=jauneF](3,0)(3.5,1.5)(6,2)(3.5,2.5)(3,4)(2.5,2.5)(0,2)(2.5,1.5)(3,0)
}

\put(19,7){
\pspolygon[fillstyle=solid,fillcolor=yellow,linecolor=jauneF](3,0)(3.5,1.5)(6,2)(3.5,2.5)(3,4)(2.5,2.5)(0,2)(2.5,1.5)(3,0)
}

\put(50,-15){
\pspolygon[fillstyle=solid,fillcolor=yellow,linecolor=jauneF](3,0)(3.5,1.5)(6,2)(3.5,2.5)(3,4)(2.5,2.5)(0,2)(2.5,1.5)(3,0)
\put(8.5,1){\footnotesize OPNET times}
\psframe[linecolor=gray,linewidth=.5pt,linestyle=dashed](0,-1.5)(30,5)
}

}

\end{pspicture}
}
\caption{Petri Net structure of the operation: cutting $1$}\label{Fig:RdP_Cutting1}
\end{figure}
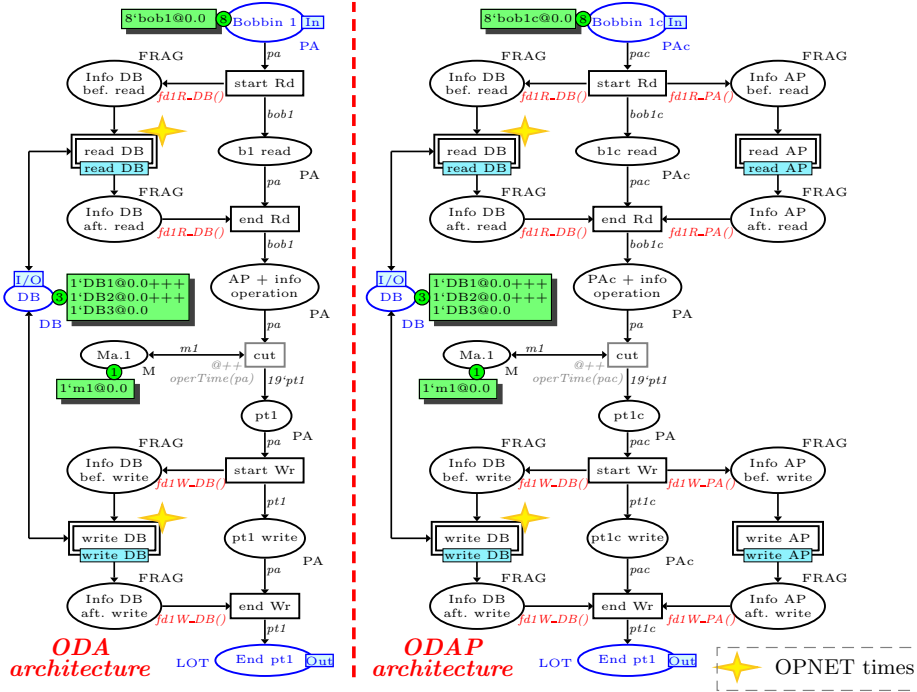

As illustrated on the evaluation architecture (figure~\ref{Fig:Syst_Eval}), the estimated \emph{round trip times} are injected into CPN Tools and more exactly, they shall be set on timed transitions which reflect the read/write actions on databases. Second views describe the considered operations. For instance, the second view of the cutting $1$ operation (introduced section~\ref{Sec3:StudyCase:adaptation}) is presented in the figure~\ref{Fig:RdP_Cutting1}. Both views of this operation, related to the ODA and ODAP architectures, are shown in the same figure because they are almost similar except for the read and write phases which will be discussed in more detail below. Three places from these two Petri Nets are bound to the first view (see figure~\ref{Fig:hier_vue1}), namely the place $Bobbin 1(c)$ (port: \emph{In}), the place $DB$ in which the three databases are defined (port: \emph{I/O} since they are shared resources between the three operations) and the place $pt1(c)$ in which cut pieces are stored (port: $Out$).

Let us focus now on the Petri Net structure of the cutting $1$ operation (see figure~\ref{Fig:RdP_Cutting1}): when a bobbin$1/c$ (i.e bob$1/c$) arrives into the queue for being cut (i.e places \emph{Bobbin 1/Bobbin 1c}), we generate straight away the data fragments needed by the machine $1$ for starting the operation (fabrication orders\ldots). Let us remind the machine $1$ needs to read the following fragments: $F1$, $F4$, $F5$ and $F8$. Considering the ODA architecture, these fragments are all allocated on the fixed databases, i.e into the place \emph{Info DB bef. read}. %(via the function \emph{fd1\_RDB()}$=1`f1++1`f4++1`f5++1`f8$).
Considering now the ODAP architecture, one part of these fragments can be allocated on the product and another part on databases. To do this, we add to the ODAP view the place \emph{Info AP bef. read} to indicate that fragments should be read on the product rather than on databases. Thus, several distribution patterns between product and fixed databases can be defined. One possible combination might be to allocate $F1$, $F5$ to the product and $F4$, $F8$ to the databases. %, expressed by the following functions in CPN Tools: \emph{fd1\_RDB()}$=1`f1++1`f5$ and \emph{fd1\_RPA()}$=1`f4++1`f8$. The total of fragments, i.e \emph{fd1\_RDB()}$+$\emph{fd1\_RPA()} is equal to the fragments defined by \emph{fd1\_RDB()} in ODA.
After having read the fragments, the cutting task (denoted "cut" transition) can start. The write phase is similar to the read phase. This Petri Net structure is used for the other operations: cutting $2$ and sewing.

Views of the third level deal with the read and write phases of fragments, either on databases and on products, i.e the transitions denoted \emph{read/write DB} and AP: figure~\ref{Fig:RdP_Cutting1}). With regard to the \emph{read/write DB} transitions, we added the statistical \emph{round trip times} extracted from OPNET taking into account the client machines, the databases and the fragments (see table~\ref{Table:tps_Opnet}). With regard to the \emph{read/write PA} transitions, we define several throughputs in the next section.

\section{Results and Analysis}\label{Sec5:Result}

%\subsection{Experimentation purpose}
Considering the ODA architecture as our reference model, the purpose of our experimentation is then to determine the factors impacting positively or negatively on the ODAP architecture performance, and to identify configurations where ODAP should be benefit. In this article, we aim to study the influences of two parameters which are the communicating product throughput and the fragment distribution pattern. %, that we assumed to have an important influence on the manufacturing time.
In fact, communicating products can exchange data with their environment at a given throughput. %We clearly believe that this parameter may have a significant impact on performances.
For our experiments, we consider $4$ levels of throughput: $100Mbps$,$54Mbps$, $11Mbps$ and $1Mbps$. %(corresponding to communicating products using RFID tags with high data transfer rate).
A fragment distribution pattern indicates the simulator how to place the different fragments, either on the distributed database or on the product as explained in the previous section. It is composed of eight boolean values [$F1$, $F2$\ldots $F8$]; $\overline{Fi}$ meaning that the fragment \emph{i} is located on the database and $Fi$ meaning the fragment \emph{i} is located on the product. For example [$\overline{F1}~\overline{F2}~\overline{F3}~\overline{F4}~\overline{F5}~F6~\overline{F7}~F8$] informs the simulator that only fragments $F8$ and $F6$ should be placed on products and the others let on the database. In practice, all the different possible dissemination patterns are tested for a given throughput, which leads to $256$ ($2^{8}$) experiments per throughput. Each experiment is simulated $10$ times and the mean times needed to produce $85$ headdresses using ODA and ODAP architectures are recorded. This number has been defined arbitrarily.

%\subsection{Experimentation results}
Table~\ref{Table:Tps_CPN} summarizes the results obtained during the experimentation. Each line of the table corresponds to a given throughput value and each column to a specific configuration of the data distribution pattern: ODA (all fragments are allocated to the databases), full ODAP (all fragments are located on the product), and best hybrid ODAP configuration (some fragments are on the database, others on the product). Time values obtained for a given throughput and configuration are then reported in the table.
As can be seen, for our scenario, full ODAP and ODA are quite similar in terms of performance, when considering $100Mbps$, $54Mps$ and $11Mbps$ throughputs. Therefore, it might appear that disseminating information all over the different informational vectors has no influence on the manufacturing time if the product throughput is high enough. With a correct throughput, it is then possible to imagine an information system completely distributed on a product network. When decreasing, the throughput yet acts as a very important constraint and full ODAP is clearly a bad solution. However, the best hybrid configuration always gives good results, which means some data can be stored on the product no matter what the throughput is.

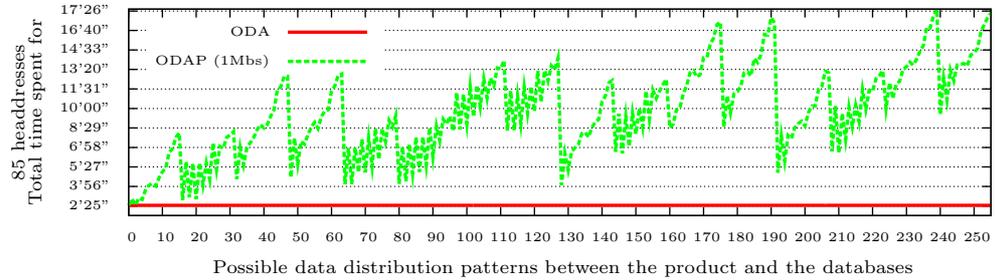
\begin{figure}[ht]
\scalebox{1}{
\centering
%\tiny{\input{PAtimes}}
\tiny{\input{PAtimes_sansMem}}
}
\caption{Comparison between ODA and ODAP (product throughput$=1Mbps$)}
\label{PA_times}
\end{figure}

\begin{figure}[ht]
\scalebox{.94}{
\centering
\scriptsize{\input{coef}}
}
\caption{Significant factors and their respective coefficient values}
\label{Fig:coef}
\end{figure}
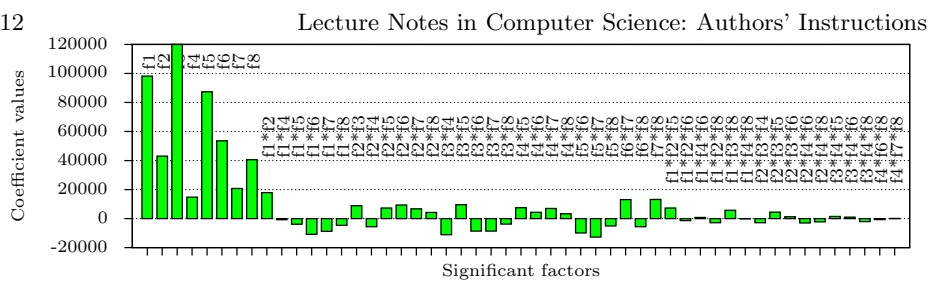

In the figure~\ref{PA_times} are plotted two curves representing the ODA and ODAP times ($y$-axis) for producing $85$ headdresses, with a throughput of $1Mb/s$. On the $x$-axis are represented the $2^{8}$ possible combinations of distribution. Clearly, the distribution pattern has a very important effect in that case. In fact, the time needed to complete the production varies from $2'25''$ to $17'26''$. %As a result, %when the throughput becomes small, it is then really important to know which pattern to use in order to prevent performance loss.
Based on these observations, it might be interesting to determine if each fragment has the same impact on the manufacturing time and if not, to identify the important fragments and evaluate their influence. To do so, a statistical analysis of the experiments done with a $1Mbps$ throughput has been initiated and showed that all fragments are important, but some of their interactions as well. An interaction between $2$ fragments means that the impact of these fragments all together on the product is different from the sum of the impacts of each fragment. In our case, interactions up to level $3$ (interactions of three fragments) have to be taken into account. In the following, we use the term "factors" to denote significant fragments and interactions. For our scenario, there are up to $51$ factors, as reported on the histogram $x$-axis in the  figure~\ref{Fig:coef}. The impact of each factor is then estimated thanks to a multiple regression analysis that determines for each factor a coefficient, related to the linear regression equation. The value of a factor coefficient could be roughly considered as the effect of this factor on the manufacturing time. The higher the coefficient value, the more the manufacturing time increases. The figure~\ref{Fig:coef} clearly shows that some fragments have a very important effect ($F1$, $F3$, $F5$), and others have a very moderate one, sometimes equal to interactions of level $2$ (e.g the effect of $F4$ is almost the same as $F7*F8$). As a result, our study demonstrates there are lots of factors with different effects to take into account when setting up an ODAP. %Besides, the effect and significance of each factor is certainly related to the application scenario, to the update frequency and the size of the factor. As a conclusion, choosing a correct dissemination pattern for a given scenario is far from being simple and is function of numerous parameters.

\vspace{-.25cm}
\begin{table}[htb]
  \caption{Times obtained for $3$ fragment distributions according to $4$ throughputs}
  \label{Table:Tps_CPN}
  \begin{center}
    \begin{tabular}{p{1.5cm} p{2cm} p{2.5cm} p{3.5cm}}
    \hline
           \centering \scriptsize Product throughput & \centering \scriptsize ODA distribution & \centering \scriptsize Full ODAP distribution & \scriptsize Best hybrid distribution \\
         \hline
         \centering \scriptsize $100Mbps$ & \centering \scriptsize $2'27''$ & \centering \scriptsize $2'28''$ & \scriptsize $2'27''$ ($F5$,$F7$)\\
         \centering \scriptsize $54Mbps$ & \centering \scriptsize $2'27''$ & \centering \scriptsize $2'28''$ & \scriptsize $2'27''$ ($F5$,$F6$,$F7$,$F8$)\\
         \centering \scriptsize $11Mbps$ & \centering \scriptsize $2'27''$ & \centering \scriptsize $2'52''$ & \scriptsize $2'27''$ ($F7$)\\
         \centering \scriptsize $1Mbps$ & \centering \scriptsize $2'27''$ & \centering \scriptsize $17'26''$ & \scriptsize $2'28''$ ($F7$)\\
         \hline
    \end{tabular}
  \end{center}
\end{table}

\vspace{-1.6cm}
\vspace{-.7cm}

\section{conclusion}
A multitude of informational vectors take place in Supply Chain environments as fixed databases or manufactured products on which we are able to embed significant proportion of data. By considering distributed database systems, specific data fragments can be embedded/allocated on these products (for example, data useful for their life cycle). The paper analyzes three specific distribution patterns to determine the different parameters impacting on the manufacturing system performance and shows that choosing a good pattern is complex. In a further work, we are willing to implement an experimental design leading us to control the best way to disseminate information on the informational vectors.

\bibliographystyle{splncs03}
\bibliography{MATICOM}

\end{document}

%% file: PAtimes_sansMem.tex
% GNUPLOT: LaTeX picture with Postscript
\begingroup
  \makeatletter
  \providecommand\color[2][]{%
    \GenericError{(gnuplot) \space\space\space\@spaces}{%
      Package color not loaded in conjunction with
      terminal option `colourtext'%
    }{See the gnuplot documentation for explanation.%
    }{Either use 'blacktext' in gnuplot or load the package
      color.sty in LaTeX.}%
    \renewcommand\color[2][]{}%
  }%
  \providecommand\includegraphics[2][]{%
    \GenericError{(gnuplot) \space\space\space\@spaces}{%
      Package graphicx or graphics not loaded%
    }{See the gnuplot documentation for explanation.%
    }{The gnuplot epslatex terminal needs graphicx.sty or graphics.sty.}%
    \renewcommand\includegraphics[2][]{}%
  }%
  \providecommand\rotatebox[2]{#2}%
  \@ifundefined{ifGPcolor}{%
    \newif\ifGPcolor
    \GPcolortrue
  }{}%
  \@ifundefined{ifGPblacktext}{%
    \newif\ifGPblacktext
    \GPblacktexttrue
  }{}%
  % define a \g@addto@macro without @ in the name:
  \let\gplgaddtomacro\g@addto@macro
  % define empty templates for all commands taking text:
  \gdef\gplbacktext{}%
  \gdef\gplfronttext{}%
  \makeatother
  \ifGPblacktext
    % no textcolor at all
    \def\colorrgb#1{}%
    \def\colorgray#1{}%
  \else
    % gray or color?
    \ifGPcolor
      \def\colorrgb#1{\color[rgb]{#1}}%
      \def\colorgray#1{\color[gray]{#1}}%
      \expandafter\def\csname LTw\endcsname{\color{white}}%
      \expandafter\def\csname LTb\endcsname{\color{black}}%
      \expandafter\def\csname LTa\endcsname{\color{black}}%
      \expandafter\def\csname LT0\endcsname{\color[rgb]{1,0,0}}%
      \expandafter\def\csname LT1\endcsname{\color[rgb]{0,1,0}}%
      \expandafter\def\csname LT2\endcsname{\color[rgb]{0,0,1}}%
      \expandafter\def\csname LT3\endcsname{\color[rgb]{1,0,1}}%
      \expandafter\def\csname LT4\endcsname{\color[rgb]{0,1,1}}%
      \expandafter\def\csname LT5\endcsname{\color[rgb]{1,1,0}}%
      \expandafter\def\csname LT6\endcsname{\color[rgb]{0,0,0}}%
      \expandafter\def\csname LT7\endcsname{\color[rgb]{1,0.3,0}}%
      \expandafter\def\csname LT8\endcsname{\color[rgb]{0.5,0.5,0.5}}%
    \else
      % gray
      \def\colorrgb#1{\color{black}}%
      \def\colorgray#1{\color[gray]{#1}}%
      \expandafter\def\csname LTw\endcsname{\color{white}}%
      \expandafter\def\csname LTb\endcsname{\color{black}}%
      \expandafter\def\csname LTa\endcsname{\color{black}}%
      \expandafter\def\csname LT0\endcsname{\color{black}}%
      \expandafter\def\csname LT1\endcsname{\color{black}}%
      \expandafter\def\csname LT2\endcsname{\color{black}}%
      \expandafter\def\csname LT3\endcsname{\color{black}}%
      \expandafter\def\csname LT4\endcsname{\color{black}}%
      \expandafter\def\csname LT5\endcsname{\color{black}}%
      \expandafter\def\csname LT6\endcsname{\color{black}}%
      \expandafter\def\csname LT7\endcsname{\color{black}}%
      \expandafter\def\csname LT8\endcsname{\color{black}}%
    \fi
  \fi
  \setlength{\unitlength}{0.0500bp}%
  \begin{picture}(7200.00,2120.00)%
  \put(500,-400){
    \gplgaddtomacro\gplbacktext{%
      \csname LTb\endcsname%
      \put(396,776){\makebox(0,0)[r]{\strut{} 2'25''}}%
      \csname LTb\endcsname%
      \put(396,923){\makebox(0,0)[r]{\strut{} 3'56''}}%
      \csname LTb\endcsname%
      \put(396,1070){\makebox(0,0)[r]{\strut{} 5'27''}}%
      \csname LTb\endcsname%
      \put(396,1216){\makebox(0,0)[r]{\strut{} 6'58''}}%
      \csname LTb\endcsname%
      \put(396,1363){\makebox(0,0)[r]{\strut{} 8'29''}}%
      \csname LTb\endcsname%
      \put(396,1509){\makebox(0,0)[r]{\strut{} 10'00''}}%
      \csname LTb\endcsname%
      \put(396,1656){\makebox(0,0)[r]{\strut{} 11'31''}}%
      \csname LTb\endcsname%
      \put(396,1803){\makebox(0,0)[r]{\strut{} 13'20''}}%
      \csname LTb\endcsname%
      \put(396,1949){\makebox(0,0)[r]{\strut{} 14'33''}}%
      \csname LTb\endcsname%
      \put(396,2096){\makebox(0,0)[r]{\strut{} 16'40''}}%
      \csname LTb\endcsname%
      \put(396,2242){\makebox(0,0)[r]{\strut{} 17'26''}}%
      \put(0,50){
      \put(528,484){\makebox(0,0){\strut{} 0}}%
      \put(780,484){\makebox(0,0){\strut{} 10}}%
      \put(1032,484){\makebox(0,0){\strut{} 20}}%
      \put(1284,484){\makebox(0,0){\strut{} 30}}%
      \put(1537,484){\makebox(0,0){\strut{} 40}}%
      \put(1789,484){\makebox(0,0){\strut{} 50}}%
      \put(2041,484){\makebox(0,0){\strut{} 60}}%
      \put(2293,484){\makebox(0,0){\strut{} 70}}%
      \put(2545,484){\makebox(0,0){\strut{} 80}}%
      \put(2797,484){\makebox(0,0){\strut{} 90}}%
      \put(3050,484){\makebox(0,0){\strut{} 100}}%
      \put(3302,484){\makebox(0,0){\strut{} 110}}%
      \put(3554,484){\makebox(0,0){\strut{} 120}}%
      \put(3806,484){\makebox(0,0){\strut{} 130}}%
      \put(4058,484){\makebox(0,0){\strut{} 140}}%
      \put(4310,484){\makebox(0,0){\strut{} 150}}%
      \put(4563,484){\makebox(0,0){\strut{} 160}}%
      \put(4815,484){\makebox(0,0){\strut{} 170}}%
      \put(5067,484){\makebox(0,0){\strut{} 180}}%
      \put(5319,484){\makebox(0,0){\strut{} 190}}%
      \put(5571,484){\makebox(0,0){\strut{} 200}}%
      \put(5823,484){\makebox(0,0){\strut{} 210}}%
      \put(6075,484){\makebox(0,0){\strut{} 220}}%
      \put(6328,484){\makebox(0,0){\strut{} 230}}%
      \put(6580,484){\makebox(0,0){\strut{} 240}}%
      \put(6832,484){\makebox(0,0){\strut{} 250}}%
      }
      \put(-150,1480){\rotatebox{90}{\makebox(0,0){\strut{}\scriptsize Total time spent for}}}%
      \put(-280,1480){\rotatebox{90}{\makebox(0,0){\strut{}\scriptsize $85$ headdresses}}}%
      \put(3743,300){\makebox(0,0){\strut{} \scriptsize Possible data distribution patterns between the product and the databases}}%
    }%
    \gplgaddtomacro\gplfronttext{%
      \csname LTb\endcsname%
      \put(1584,2083){\makebox(0,0)[r]{\strut{}\tiny ODA}}%
      \csname LTb\endcsname%
      \put(1584,1863){\makebox(0,0)[r]{\strut{}\tiny ODAP (1Mbs)}}%
    }%
    \gplbacktext
    \put(0,0){\includegraphics{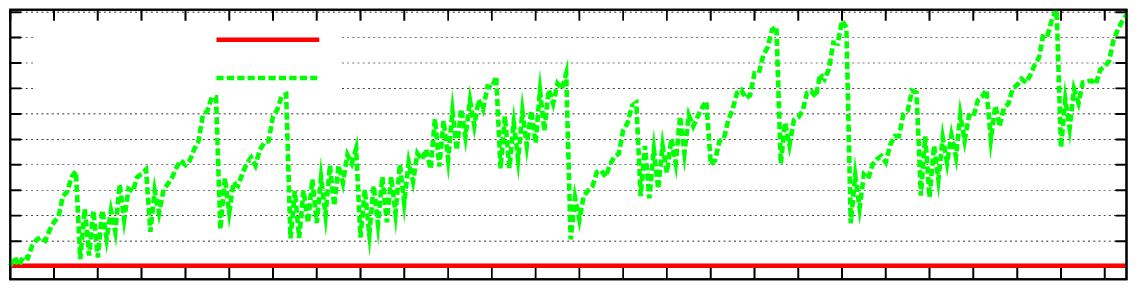}}%
    \gplfronttext
    }
  \end{picture}%
\endgroup

%% file: coef.tex
% GNUPLOT: LaTeX picture with Postscript
\begingroup
  \makeatletter
  \providecommand\color[2][]{%
    \GenericError{(gnuplot) \space\space\space\@spaces}{%
      Package color not loaded in conjunction with
      terminal option `colourtext'%
    }{See the gnuplot documentation for explanation.%
    }{Either use 'blacktext' in gnuplot or load the package
      color.sty in LaTeX.}%
    \renewcommand\color[2][]{}%
  }%
  \providecommand\includegraphics[2][]{%
    \GenericError{(gnuplot) \space\space\space\@spaces}{%
      Package graphicx or graphics not loaded%
    }{See the gnuplot documentation for explanation.%
    }{The gnuplot epslatex terminal needs graphicx.sty or graphics.sty.}%
    \renewcommand\includegraphics[2][]{}%
  }%
  \providecommand\rotatebox[2]{#2}%
  \@ifundefined{ifGPcolor}{%
    \newif\ifGPcolor
    \GPcolortrue
  }{}%
  \@ifundefined{ifGPblacktext}{%
    \newif\ifGPblacktext
    \GPblacktexttrue
  }{}%
  % define a \g@addto@macro without @ in the name:
  \let\gplgaddtomacro\g@addto@macro
  % define empty templates for all commands taking text:
  \gdef\gplbacktext{}%
  \gdef\gplfronttext{}%
  \makeatother
  \ifGPblacktext
    % no textcolor at all
    \def\colorrgb#1{}%
    \def\colorgray#1{}%
  \else
    % gray or color?
    \ifGPcolor
      \def\colorrgb#1{\color[rgb]{#1}}%
      \def\colorgray#1{\color[gray]{#1}}%
      \expandafter\def\csname LTw\endcsname{\color{white}}%
      \expandafter\def\csname LTb\endcsname{\color{black}}%
      \expandafter\def\csname LTa\endcsname{\color{black}}%
      \expandafter\def\csname LT0\endcsname{\color[rgb]{1,0,0}}%
      \expandafter\def\csname LT1\endcsname{\color[rgb]{0,1,0}}%
      \expandafter\def\csname LT2\endcsname{\color[rgb]{0,0,1}}%
      \expandafter\def\csname LT3\endcsname{\color[rgb]{1,0,1}}%
      \expandafter\def\csname LT4\endcsname{\color[rgb]{0,1,1}}%
      \expandafter\def\csname LT5\endcsname{\color[rgb]{1,1,0}}%
      \expandafter\def\csname LT6\endcsname{\color[rgb]{0,0,0}}%
      \expandafter\def\csname LT7\endcsname{\color[rgb]{1,0.3,0}}%
      \expandafter\def\csname LT8\endcsname{\color[rgb]{0.5,0.5,0.5}}%
    \else
      % gray
      \def\colorrgb#1{\color{black}}%
      \def\colorgray#1{\color[gray]{#1}}%
      \expandafter\def\csname LTw\endcsname{\color{white}}%
      \expandafter\def\csname LTb\endcsname{\color{black}}%
      \expandafter\def\csname LTa\endcsname{\color{black}}%
      \expandafter\def\csname LT0\endcsname{\color{black}}%
      \expandafter\def\csname LT1\endcsname{\color{black}}%
      \expandafter\def\csname LT2\endcsname{\color{black}}%
      \expandafter\def\csname LT3\endcsname{\color{black}}%
      \expandafter\def\csname LT4\endcsname{\color{black}}%
      \expandafter\def\csname LT5\endcsname{\color{black}}%
      \expandafter\def\csname LT6\endcsname{\color{black}}%
      \expandafter\def\csname LT7\endcsname{\color{black}}%
      \expandafter\def\csname LT8\endcsname{\color{black}}%
    \fi
  \fi
  \setlength{\unitlength}{0.0500bp}%
  \begin{picture}(7200.00,1500.00)%
  \put(200,-2300){
    \gplgaddtomacro\gplbacktext{%
      \csname LTb\endcsname%
      \put(597,2395){\makebox(0,0)[r]{\strut{}-20000}}%
      \csname LTb\endcsname%
      \put(597,2627){\makebox(0,0)[r]{\strut{} 0}}%
      \csname LTb\endcsname%
      \put(597,2859){\makebox(0,0)[r]{\strut{} 20000}}%
      \csname LTb\endcsname%
      \put(597,3091){\makebox(0,0)[r]{\strut{} 40000}}%
      \csname LTb\endcsname%
      \put(597,3324){\makebox(0,0)[r]{\strut{} 60000}}%
      \csname LTb\endcsname%
      \put(597,3556){\makebox(0,0)[r]{\strut{} 80000}}%
      \csname LTb\endcsname%
      \put(597,3788){\makebox(0,0)[r]{\strut{} 100000}}%
      \csname LTb\endcsname%
      \put(597,4020){\makebox(0,0)[r]{\strut{} 120000}}%
      \put(75,650){
      \put(865,3300){\rotatebox{90}{\makebox(0,0)[r]{\strut{}f1}}}%
      \put(983,3300){\rotatebox{90}{\makebox(0,0)[r]{\strut{}f2}}}%
      \put(1102,3300){\rotatebox{90}{\makebox(0,0)[r]{\strut{}f3}}}%
      \put(1220,3300){\rotatebox{90}{\makebox(0,0)[r]{\strut{}f4}}}%
      \put(1339,3300){\rotatebox{90}{\makebox(0,0)[r]{\strut{}f5}}}%
      \put(1457,3300){\rotatebox{90}{\makebox(0,0)[r]{\strut{}f6}}}%
      \put(1576,3300){\rotatebox{90}{\makebox(0,0)[r]{\strut{}f7}}}%
      \put(1695,3300){\rotatebox{90}{\makebox(0,0)[r]{\strut{}f8}}}%
      }
      \put(75,150){
      \put(1813,3300){\rotatebox{90}{\makebox(0,0)[r]{\strut{}f1*f2}}}%
      \put(1932,3300){\rotatebox{90}{\makebox(0,0)[r]{\strut{}f1*f4}}}%
      \put(2050,3300){\rotatebox{90}{\makebox(0,0)[r]{\strut{}f1*f5}}}%
      \put(2169,3300){\rotatebox{90}{\makebox(0,0)[r]{\strut{}f1*f6}}}%
      \put(2288,3300){\rotatebox{90}{\makebox(0,0)[r]{\strut{}f1*f7}}}%
      \put(2406,3300){\rotatebox{90}{\makebox(0,0)[r]{\strut{}f1*f8}}}%
      \put(2525,3300){\rotatebox{90}{\makebox(0,0)[r]{\strut{}f2*f3}}}%
      \put(2643,3300){\rotatebox{90}{\makebox(0,0)[r]{\strut{}f2*f4}}}%
      \put(2762,3300){\rotatebox{90}{\makebox(0,0)[r]{\strut{}f2*f5}}}%
      \put(2880,3300){\rotatebox{90}{\makebox(0,0)[r]{\strut{}f2*f6}}}%
      \put(2999,3300){\rotatebox{90}{\makebox(0,0)[r]{\strut{}f2*f7}}}%
      \put(3118,3300){\rotatebox{90}{\makebox(0,0)[r]{\strut{}f2*f8}}}%
      \put(3236,3300){\rotatebox{90}{\makebox(0,0)[r]{\strut{}f3*f4}}}%
      \put(3355,3300){\rotatebox{90}{\makebox(0,0)[r]{\strut{}f3*f5}}}%
      \put(3473,3300){\rotatebox{90}{\makebox(0,0)[r]{\strut{}f3*f6}}}%
      \put(3592,3300){\rotatebox{90}{\makebox(0,0)[r]{\strut{}f3*f7}}}%
      \put(3710,3300){\rotatebox{90}{\makebox(0,0)[r]{\strut{}f3*f8}}}%
      \put(3829,3300){\rotatebox{90}{\makebox(0,0)[r]{\strut{}f4*f5}}}%
      \put(3948,3300){\rotatebox{90}{\makebox(0,0)[r]{\strut{}f4*f6}}}%
      \put(4066,3300){\rotatebox{90}{\makebox(0,0)[r]{\strut{}f4*f7}}}%
      \put(4185,3300){\rotatebox{90}{\makebox(0,0)[r]{\strut{}f4*f8}}}%
      \put(4303,3300){\rotatebox{90}{\makebox(0,0)[r]{\strut{}f5*f6}}}%
      \put(4422,3300){\rotatebox{90}{\makebox(0,0)[r]{\strut{}f5*f7}}}%
      \put(4540,3300){\rotatebox{90}{\makebox(0,0)[r]{\strut{}f5*f8}}}%
      \put(4659,3300){\rotatebox{90}{\makebox(0,0)[r]{\strut{}f6*f7}}}%
      \put(4778,3300){\rotatebox{90}{\makebox(0,0)[r]{\strut{}f6*f8}}}%
      \put(4896,3300){\rotatebox{90}{\makebox(0,0)[r]{\strut{}f7*f8}}}%
      \put(5015,3300){\rotatebox{90}{\makebox(0,0)[r]{\strut{}f1*f2*f5}}}%
      \put(5133,3300){\rotatebox{90}{\makebox(0,0)[r]{\strut{}f1*f2*f6}}}%
      \put(5252,3300){\rotatebox{90}{\makebox(0,0)[r]{\strut{}f1*f4*f6}}}%
      \put(5371,3300){\rotatebox{90}{\makebox(0,0)[r]{\strut{}f1*f2*f8}}}%
      \put(5489,3300){\rotatebox{90}{\makebox(0,0)[r]{\strut{}f1*f3*f8}}}%
      \put(5608,3300){\rotatebox{90}{\makebox(0,0)[r]{\strut{}f1*f4*f8}}}%
      \put(5726,3300){\rotatebox{90}{\makebox(0,0)[r]{\strut{}f2*f3*f4}}}%
      \put(5845,3300){\rotatebox{90}{\makebox(0,0)[r]{\strut{}f2*f3*f5}}}%
      \put(5963,3300){\rotatebox{90}{\makebox(0,0)[r]{\strut{}f2*f3*f6}}}%
      \put(6082,3300){\rotatebox{90}{\makebox(0,0)[r]{\strut{}f2*f4*f6}}}%
      \put(6201,3300){\rotatebox{90}{\makebox(0,0)[r]{\strut{}f2*f4*f8}}}%
      \put(6319,3300){\rotatebox{90}{\makebox(0,0)[r]{\strut{}f3*f4*f5}}}%
      \put(6438,3300){\rotatebox{90}{\makebox(0,0)[r]{\strut{}f3*f4*f6}}}%
      \put(6556,3300){\rotatebox{90}{\makebox(0,0)[r]{\strut{}f3*f4*f8}}}%
      \put(6675,3300){\rotatebox{90}{\makebox(0,0)[r]{\strut{}f4*f6*f8}}}%
      \put(6793,3300){\rotatebox{90}{\makebox(0,0)[r]{\strut{}f4*f7*f8}}}%
      }
      \put(-100,3207){\rotatebox{90}{\makebox(0,0){\strut{}Coefficient values}}}%
      \put(3875,2200){\makebox(0,0){\strut{}Significant factors}}%
    }%
    \gplgaddtomacro\gplfronttext{%
    }%
    \gplbacktext
    \put(0,0){\includegraphics{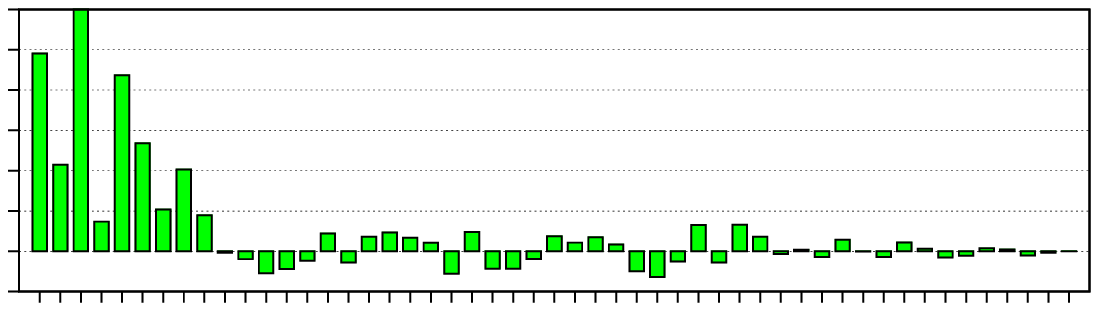}}%
    \gplfronttext
    }
  \end{picture}%
\endgroup